\documentclass[journal]{IEEEtran}
\usepackage{amsfonts}
\usepackage[cmex10]{amsmath}
\usepackage{cite}
\usepackage{array}
\usepackage{mdwmath}
\usepackage{mdwtab}
\usepackage{graphicx,cite,epsfig}
\usepackage{eqparbox}
\usepackage{subfigure}
\usepackage{stfloats}
\usepackage{url}
\usepackage{amssymb}
\usepackage{float}
\usepackage{indentfirst}
\usepackage{makeidx}
\usepackage{tabularx,ragged2e}
\usepackage{cases}
\usepackage{algorithm}
\usepackage{algorithmic}
\usepackage{mathtools}
\usepackage{color}
\usepackage{bm}
\usepackage{setspace}
\usepackage{makecell}
\usepackage{xcolor}
\allowdisplaybreaks[4]

\newcommand{\tabincell}[2]{\begin{tabular}{@{}#1@{}}#2\end{tabular}}
\newcolumntype{C}{>{\Centering\arraybackslash}X}

\begin{document}

\title{Joint Active User Detection and Channel Estimation in Massive Access Systems Exploiting Reed-Muller Sequences}

\author{Jue~Wang,~\IEEEmembership{Student~Member,~IEEE,}
        Zhaoyang~Zhang,~\IEEEmembership{Member,~IEEE,}
        and Lajos Hanzo,~\IEEEmembership{Fellow,~IEEE}
        %
\thanks{This work was supported in part by National Natural Science Foundation of China under Grant 61725104 and 61631003, and Huawei Technologies Co., Ltd under Grant HF2017010003, YB2015040053 and YB2013120029.
}
\thanks{Jue~Wang (e-mail:
juew@zju.edu.cn) and Zhaoyang~Zhang (Corresponding Author, e-mail: zhzy@zju.edu.cn) are with the College of Information Science and Electronic
Engineering, Zhejiang University, Hangzhou 310027, China. Lajos~Hanzo (e-mail: lh@ecs.soton.ac.uk) is with the Department of Electronics and Computer Science, University of Southampton, UK. 
}}%

\maketitle
\vspace{-1cm}
\begin{abstract}
The requirements to support massive connectivity and low latency in
massive Machine Type Communications (mMTC) bring a huge challenge
in the design of its random access (RA) procedure, which usually calls
for efficient joint active user detection and channel estimation.
In this paper, we exploit the vast sequence space and the beneficial nested structure of the length-$2^m$ second-order Reed-Muller
(RM) sequences for designing an efficient RA scheme, which is capable of reliably detecting multiple active
users from the set of unknown potential users with a size as large as $2^{m(m-1)/2}$, whilst simultaneously
estimating their channel state information as well.
Explicitly, at the transmitter each user is
mapped to a specially designed RM sequence, which facilitates reliable
joint sequence detection and channel estimation based on a single
transmission event. To elaborate, as a first step, at the receiver we
exploit the elegant nested structure of the RM sequences using a
layer-by-layer RM detection algorithm for the single-user
(single-sequence) scenario. Then an iterative RM detection and channel
estimation algorithm is conceived for the multi-user (multi-sequence)
scenario. As a benefit of the information exchange between the RM
sequence detector and channel estimator, a compelling performance
vs. complexity trade-off is struck, as evidenced both by our
analytical and numerical results.
\end{abstract}


\IEEEpeerreviewmaketitle

\section{Introduction}
\subsection{Motivation}
The 5G wireless mobile network aims to support diverse applications, which
were classified by 3GPP into three categories: enhanced Mobile
Broadband (eMBB), Ultra-Reliable and Low Latency Communications
(URLLC), and massive Machine Type Communications (mMTC)\cite{5G}. For
the mMTC scenario, the key requirement is to provide massive
connectivity for instant access of short
and sporadic data traffic in applications like smart city and industrial internet of things \cite{5G_Huawei}.

\begin{figure}[!htp]
\small
\centering
\includegraphics[width=0.45\textwidth]{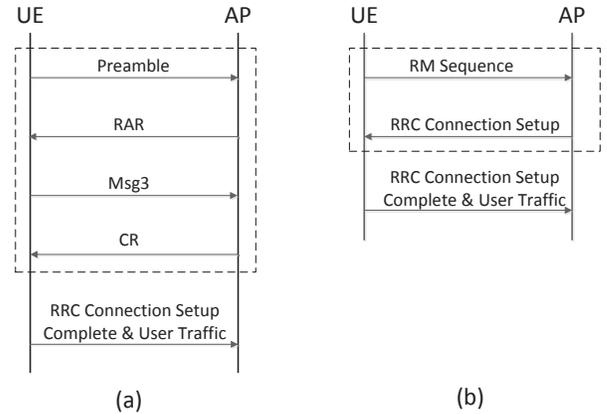}
\caption{(a) The RA procedure of LTE system. (b) The RA procedure using RM sequences.}
\label{RA_procedure}
\end{figure}

The random access (RA) procedure is key to enable the massive connectivity. As depicted in Fig. \ref{RA_procedure} (a), in the current long term evolution (LTE) system, a four-message
handshake between the user equipment (UE) and the access point (AP),
which consists of Preamble, Random Access Response (RAR), Initial Layer 3
Message (Msg3) and Contention Resolution (CR)
\cite{four_message_handshake}, is used for RA. If the UE successfully accesses the
system, it completes the RA procedure by sending the ``RRC Connection
Setup Complete" message and starts to transmit its data. When applied
to mMTC, this RA scheme faces two challenges.  Firstly, the
small-sized sequence space suffers from a high collision rate.  When a
UE triggers the RA procedure, it randomly chooses a preamble from the
available sequence pool to send in the next RA slot.  According to LTE
\cite{ZC}, there are 54 Zadoff-Chu (ZC) sequences reserved for
contention in each cell. If more than one UE select the same
sequence, a collision occurs and these UEs must retransmit the
sequence.  In the context of massive connectivity, the frequent
collisions and precipitated retransmissions will lead to network
congestion, increased delay and resource wastage.  Secondly, the traffic pattern exchanged in machine-type communications is typically
short and sporadic \cite{short_packet}.
A large amount of signalling will lead to high signalling-to-data ratio and low bandwidth efficiency.
Since the ZC sequences do not carry any information about user identifiers (IDs) in the LTE system, the AP has no knowledge of exactly who is requiring access until Msg3 has been successfully received.

Hence, the preamble sequences play a vital role in the RA procedure.
To satisfy the tight specifications of mMTC, the sequence
space has to match the number of users in the system.  Furthermore,
predefined sequences with user IDs embedded in them are preferred for the sake of low signalling overhead and system latency.  In this case,
the AP has to detect the IDs of the active users and estimate the channel
coefficients based on the received signal. To reduce congestion and
achieve low latency, the above procedures have to be accomplished with
high accuracy and at low complexity.

Moreover, it is worth pointing out that the difficulty of detecting active users from sets
of potential users with different set sizes is very different. In general, detecting active users from a
set of tens or hundreds of potential users can be readily achieved by elaborate sequence design and
exhaustive search. However, it is rather challenging to do this in the massive access scenario
where the number of unknown potential users can be as huge as hundreds of thousands or even millions,
which makes the detection reminiscent of
looking for the needle in a haystack.

\subsection{Related Work}
Numerous researchers have improved the RA procedures by designing
beneficial preamble sequences.  Some of them focus on increasing the
number of available ZC sequences. For example, the authors of \cite{preamble_reuse} proposed a preamble reuse scheme by partitioning the cell coverage and reducing the cyclic shift size.  Apart from the classic ZC sequences,
\cite{insert} they inserted extra auxiliary sequences in the positions
randomly chosen from the first data sub-frame.  However, the use of
auxiliary sequences decreases the data transmission efficiency and
increases the detection complexity.  By contrast, the RA procedure
of~\cite{code_expand} treats several RA slots as a virtual frame and
the active users randomly choose a ZC sequence for their transmission
in each slot.  This is equivalent to expanding the sequence space at
the cost of an increased overhead ratio and detection complexity.

Other authors investigate different sequence generation methods in
support of massive connectivity and active user detection.  On one
hand, considering the huge number of potential users in the system, it
is impossible to assign orthogonal sequences to each user.  On the
other hand, non-orthogonal sequences would inevitably impose
multi-user interference during the detection.  Hence, the key
challenge is to obtain a relatively large sequence space, whilst
simultaneously attaining reliable detection.
Based on the sparsity of user-activity caused by the sporadic traffic in mMTC, the active user
detection is often formulated as a compressive sensing (CS) problem.
The active users transmit a unique sequence to access the system.  To
satisfy the so-called Restricted Isometry Property (RIP) of
CS~\cite{RIP}, the entries of these sequences are usually generated
either by Gaussian or Bernoulli processes~\cite{AMP_3}.
Hence, considerable storage resources are required for storing these
sequences at the AP.  At the receiver, CS algorithms, such as Basis
Pursuit (BP) \cite{basis_pursuit}, Orthogonal Matching Pursuit (OMP)
\cite{OMP} and approximate message passing (AMP)
\cite{AMP_1}\cite{AMP_2}, are utilized for detecting the active users.
However, all these algorithms impose a high complexity that is related
either to the size of sequence space or to the number of potential
users in the system.  Based on the code-expanded scheme of
\cite{code_expand}, the sequences constructed by Bloom filtering are
utilized in \cite{bloom_filtering}.  Despite the lower latency
compared to LTE, the hash functions used during the sequence
generation have to be communicated both to all the users and to the
AP, which imposes extra signalling overhead.

In \cite{RM_as_pilot}, second-order RM sequences are invoked for
massive access in 5G.  Compared to ZC sequences, RM sequences have the
following distinct advantages:
\begin{itemize}
  \item The size of RM sequence space is several orders of magnitudes
    larger than that of the same-length ZC sequences, which meets the
    requirement of massive connectivity.
  \item The RM sequences can be uniquely and unambiguously mapped to user IDs. Hence, the AP can immediately infer the user ID right
    after the RM sequences have been correctly detected. Thus, the
    four-message handshake depicted in Fig. \ref{RA_procedure}(a) is
    simplified to the twin-step interaction shown in
    Fig. \ref{RA_procedure}(b), which results both in reduced
    signalling overhead and in reduced system latency.
  \item The structural properties of RM sequences can be exploited to
    design efficient detection algorithms having a low computational
    complexity.
\end{itemize}

In \cite{generation_of_RM}, RM sequences are used to construct the
deterministic measurement matrix of CS, where an efficient
reconstruction algorithm is also conceived, whose complexity is
determined by the length of RM sequences and by the sparsity factor
instead of the sequence space cardinality.  However, this algorithm
suffers from a serious performance degradation when the sparsity
factor increases.  Based on the algorithm in
\cite{generation_of_RM}, a beneficial shuffle operation is advocated
in \cite{RM_as_pilot} for improving the performance of RM detection,
albeit at a high computational complexity. However, the compelling
structural properties of RM sequences have not been fully exploited.
In addition, to cope with the multi-sequence-induced interference, traditional successive interference cancellation (SIC)
is adopted both in \cite{generation_of_RM} and in \cite{RM_as_pilot},
where the strongest signal is detected first and then subtracted from the received signal before detecting the next signal,
regardless of whether the detected sequence is correct or not. Hence,
once the sequence is incorrectly detected, it will affect the
performance of all subsequent detection stages.

\subsection{Main Contributions}
Considering the above attractive merits of RM sequences, we exploit
them to facilitate active user detection and channel estimation in
massive access systems in this paper.  At the transmitter, each active
user maps its ID to a unique RM sequence and then sends it to the AP.
Here, a novel mapping rule is proposed.  On one
hand, it helps to improve the sequence detection
performance.
On the other hand, it enables us to accomplish sequence detection and channel estimation at the same time, without invoking an extra channel estimation step as the algorithms in \cite{generation_of_RM} and \cite{RM_as_pilot}.

At the receiver, we firstly design a single-sequence detection
algorithm based on the nested structure of RM sequences, which reveals the relationships between its
sub-sequences.  Exploiting this structural property in the
sequence detection algorithm, we intend to improve the detection
performance at a reduced complexity.  Moreover, to cope with the
potential error propagation problem of the above algorithm, an
enhanced algorithm based on multiple lists is also conceived.  As for
the case where multiple active users coexist in the system, an
iterative RM detection and channel estimation algorithm is adopted.
At each stage of an iteration, we focus on a particular
user. Considering other signals as noise, the single-sequence
detection algorithm is invoked for updating this user's sequence
detection result.
After all the sequences are updated, the results are
fed into the channel estimator to improve the accuracy of channel
estimates.  Then the enhanced channel estimates are utilized in the
next iteration to reinforce the sequence detection.

Against the above background, our main contributions are:
\begin{itemize}
  \item A novel mapping rule between user IDs and RM sequences is
    proposed, which helps to improve the sequence detection
    capability and simplify our high-accuracy channel estimation.

  \item We derive an explicit relationship between the RM sequence and
    its sub-sequences, which is referred to as \textit{nested
      structure}. We exploit this structural property for conceiving a
    low-complexity algorithm for single sequence detection, which
    exhibits superior detection performance and channel estimation
    accuracy. In order to mitigate the error propagation, a
    moderate-complexity enhanced algorithm is also conceived.
  \item Iterative active-user detection and channel
    estimation scheme is then also proposed for the multi-sequence
    scenario. On one hand, the iterative detection may reduce the
    impact of incorrectly detected sequences. On the other hand, owing
    to the bi-directional information exchange between the active-user detector and channel estimator, the detection capability and channel estimation accuracy can mutually reinforce each other.
\end{itemize}

\subsection{Paper Organization and Notations}
The rest of the paper is organized as follows.  Section
\uppercase\expandafter{\romannumeral2} describes the system model and
introduces the transmitter and receiver structure of RA scheme using
RM sequences.  The nested structure of second-order RM sequences is
derived in Section \uppercase\expandafter{\romannumeral3}.  The blind
active user detection and channel estimation algorithms are given in
Section \uppercase\expandafter{\romannumeral4}.  Section
\uppercase\expandafter{\romannumeral5} depicts the analytical model of
the proposed algorithms and outlines the properties of them.
Simulation Results are given in Section
\uppercase\expandafter{\romannumeral6}.  And finally, Section
\uppercase\expandafter{\romannumeral7} concludes the paper.

Throughout this paper, scalars are represented in lower-case letters.
Boldface lowercase and uppercase letters denote column vectors and
matrices, respectively.  $\mathbf{0}^{N}$ and $\mathbf{1}^{N}$ are
$N\times 1$ column vectors of all zeros and all ones, respectively.
We use ${( \cdot )^T}$ to denote transpose of a matrix or vector and
${( \cdot )^*}$ to denote complex conjugate.  $\left\| \cdot \right\|$
is the Euclidean norm.  The symbol $\odot$ represents element-wise
product and $\oplus$ represents element-wise modulo-2 addition.  $x
\sim \mathcal{CN}{\left( {\mu,{\sigma ^2}} \right)}$ indicates that
$x$ is a complex Gaussian random variable with mean $\mu$ and variance
$\sigma^2$.  ${\cal O}\left( \cdot \right)$ is reserved for complexity
estimation.


\section{System Model}
The massive uplink connectivity scenario illustrated in Fig. 2 is considered, where a single AP is serving a huge pool of $C$ users and the maximum value of $C$ depends on the sequence space size.
All the users are assumed to be synchronized and each one of them has a unique user ID.
In a specific RA slot, there are $K$ active users in the system and the active user set is denoted as $\kappa\triangleq \{1,2,\ldots,K\}$.
Due to the fact that the traffic pattern in mMTC scenario is typically sporadic, it is satisfied that $K\ll C$.
\begin{figure}[!htp]
\small
\centering
\includegraphics[width=0.45\textwidth]{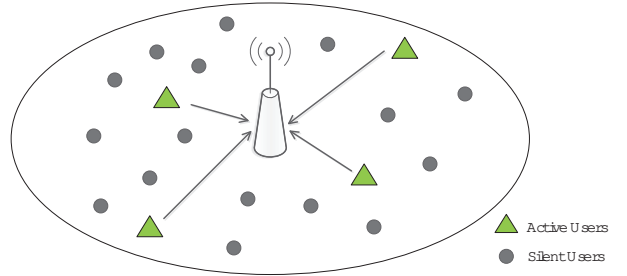}
\caption{The system model of the uplink massive connectivity scenario, where only a subset of users are active because of the sporadic traffic in mMTC.}
\label{system model}
\end{figure}

\begin{figure}[!htp]
\small
\centering
\includegraphics[width=0.45\textwidth]{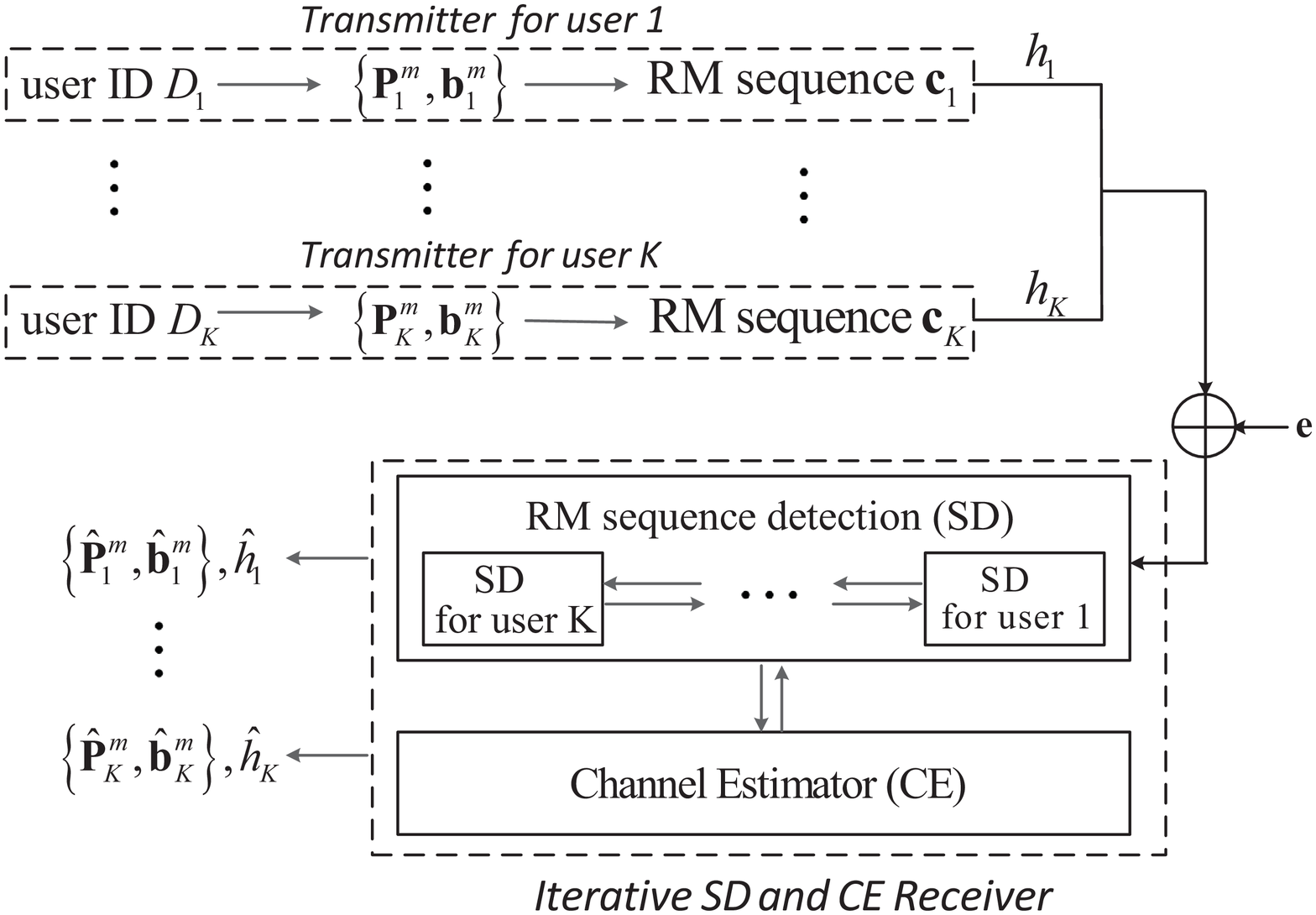}
\caption{The transmitter and receiver structure of random access scheme using RM sequences.}
\label{transmitter_and_receiver_structure}
\end{figure}

The transmitter and receiver structure is shown in
Fig. \ref{transmitter_and_receiver_structure}.  To
access the system, every active user will simultaneously send an RM
sequence along with its ID to the AP in the next available RA
slot. The AP has to accomplish both active user detection and
channel estimation based on the received signal.  In the
sequel, the processes at the transmitter and the receiver are
described in detail, respectively.

A length-$2^m$ second-order RM sequence $\mathbf{c}^m$ is determined
by an $(m\times m)$-element symmetric binary matrix $\mathbf{P}^{m}$ and an
$(m\times 1)$-element binary vector $\mathbf{b}^{m}$. According to
\cite{generation_of_RM}, given the matrix-vector pair $\left\{
     {{{\mathbf{P}}^m},{{\mathbf{b}}^m}} \right\}$, the $j$-th entry
     of the RM sequence $\mathbf{c}^m$ is expressed as
\begin{equation}\label{RMsequence_generate}
c^{m}_j=\frac{(-1)^{\mathrm{wt}(\mathbf{b}^{m})}}{\sqrt{2^{m}}}i^{(2\mathbf{b}^{m}+\mathbf{P}^{m}\mathbf{a}^{m}_{j-1})^{T}\mathbf{a}^{m}_{j-1}},j=1,\ldots,2^m,
\end{equation}
where $i$ is the imaginary unit that satisfies $i^2=-1$, $\mathrm{wt}(\mathbf{b}^{m})$ denotes the Hamming weight of the vector $\mathbf{b}^{m}$, i.e., the number of $1$s in  $\mathbf{b}^{m}$, and $\mathbf{a}^{m}_{j-1}$ is the $m$-bit binary expression of $(j-1)$.

To generate its RM sequence, the active user $k,k\!\!\in\!\!\kappa$, first maps the user ID $D_{k}\in \{0,1,\ldots,C-1\}$ to the
matrix-vector pair $\{\mathbf{P}_{k}^{m},\mathbf{b}_{k}^m\}$. We
propose a novel mapping rule as follows:
\begin{enumerate}
  \item Convert the user ID $D_{k}$ to the length-$[{m(m-1)}/{2}]$ binary vector $\left[d_{k,1},\ldots,d_{k,{m(m-1)}/{2}}\right]$.
  \item Put these bits successively into the symmetric matrix $\mathbf{P}_{k}^{m}$ having zero diagonal elements, i.e.,
\begin{equation*}
\begin{aligned}
  &\hspace*{-24pt}{\mathbf{P}}_k^m = \left[ {\begin{array}{*{10}{c}}
  0&{{p_{k,\left( {m,1} \right)}}}& \cdots &{{p_{k,\left( {m,m - 1} \right)}}} \\
  {{p_{k,\left( {m,1} \right)}}}&0& \cdots &{{p_{k,\left( {m - 1,m - 2} \right)}}} \\
   \vdots & \vdots & \ddots & \vdots  \\
  {{p_{k,\left( {m,m - 1} \right)}}}&{{p_{k,\left( {m - 1,m - 2} \right)}}}& \cdots &0
\end{array}} \right] \hfill \\
&\hspace*{-8pt}= \left[ {\begin{array}{*{20}{c}}
  0&{{d_{k,1}}}& \cdots &{{d_{k,m - 1}}} \\
  {{d_{k,1}}}&0& \cdots &{{d_{k,2m - 3}}} \\
  {{d_{k,2}}}&{{d_{k,m}}}& \cdots &{{d_{k,3m - 6}}} \\
   \vdots & \vdots & \ddots & \vdots  \\
  {{d_{k,m - 1}}}&{{d_{k,2m - 3}}}& \cdots &0
\end{array}} \right] \hfill. \\
\end{aligned}
  \end{equation*}
  \item The vector ${{\bf{b}}_k^m}=\left[{{b_{k,m}}},{{b_{k,m - 1}}},\cdots,{{b_{k,1}}} \right]^T$ is constructed according to
  \begin{equation}\label{b}
    \hspace*{-15pt}{b_{k,s}}= \left\{ \begin{gathered} {p_{k,\left( {s,1} \right)}}
      \oplus {p_{k,\left( {s,2} \right)}}\oplus \ldots \oplus
             {p_{k,\left( {s,s - 1} \right)}},\; 2 \leqslant s
             \leqslant m \hfill \\ {b_{k,m}} \oplus {b_{k,m - 1}}
             \oplus \cdots \oplus {b_{k,2}},\quad \;s = 1 \hfill \\
\end{gathered}  \right..
  \end{equation}
\end{enumerate}

Finally, the active user $k$ generates the RM sequence $\mathbf{c}^m_k$ with the mapped matrix-vector pair
$\{\mathbf{P}_{k}^{m},\mathbf{b}_{k}^m\}$.  According to the above
mapping rule, both the Hamming weight of the vector $\mathbf{b}_k^{m}$ and
the term
$(\mathbf{P}_k^{m}\mathbf{a}^{m}_{j-1})^{T}\mathbf{a}^{m}_{j-1}$ must
be even.  With the normalization coefficient $1/\sqrt{2^m}$ being omitted, the generation function shown in
(\ref{RMsequence_generate}) can be simplified to
\begin{equation}\label{RM_generate_recast}
c_{k,j}^m = {( - 1)^{{{({\mathbf{b}}_k^m)}^T}{\mathbf{a}}_{j - 1}^m + \frac{1}{2}{{({\mathbf{a}}_{j - 1}^m)}^T}{\mathbf{P}}_k^m{\mathbf{a}}_{j - 1}^m}}.
\end{equation}
The size of the sequence space created by the proposed mapping rule is up to $2^{m(m-1)/2}$ and the same-sized user space can be supported. In this sense, RM sequences are more suitable for massive access in mMTC scenario.
Furthermore, the above mapping method also brings benefits for channel estimation and sequence detection, which will be specified in Section \uppercase\expandafter{\romannumeral5}.

All the active users in this RA slot simultaneously send their RM sequences to the AP and the received signal can be expressed as
\begin{equation}\label{received signal}
{y_j^m}=\sum\limits_{k = 1}^K {{h_k}\cdot c_{k,j}^m}+ {e^m_j},\;j = 1, \ldots ,{2^m},
\end{equation}
where $h_k$ represents the channel coefficient between the active user $k$ and the AP, which is modeled as a complex Gaussian random
process with unit variance, and $e_j$ is the complex additive white
Gaussian noise (AWGN) having mean zero and variance $N_0$.  After
receiving the signal, the AP has to detect all $K$ active users and
estimate the corresponding channel coefficients.  Thanks to the one-to-one mapping between
the matrix-vector pair and the user ID, active user detection can
be accomplished by recovering the matrix-vector pairs
$\{\mathbf{P}_{k}^{m},\mathbf{b}_{k}^m\}$ from the received signal.

\section{The nested structure of RM sequences}
In this section, we introduce an important structural property of RM
sequences, which is referred to as \emph{nested structure} and is the
fundamental of the following proposed RM detection algorithms.  For
ease of exposition, several relevant definitions are given as follows.

\noindent \textbf{Definition 1}:
Given the matrix-vector pair $\{\mathbf{P}^{m},\mathbf{b}^m\}$
and the RM sequence $\mathbf{c}^{m}$ generated by them, the lower
right $(s\times{s})$-element sub-matrix of $\mathbf{P}^{m}$ is defined as its
\emph{$s$-th order sub-matrix} $\mathbf{P}^{s}$ and the vector
${\bm{\alpha }}^s \triangleq {\left[ {{p_{s,1}}, \ldots ,{p_{s,s-1}}}
    \right]^T}$ denotes the \textit{$s$-th layer }of
$\mathbf{P}^{m}$. Furthermore, the \emph{$s$-th order sub-vector} of
$\mathbf{b}^{m}$ is defined as ${{\bf{b}}^s} \buildrel \Delta \over =
\left[ {{b_s},{b_{s - 1}}, \ldots ,{b_1}} \right]^{T}$.  Then, the
matrix-vector pair can be expressed in the following recursive form
\begin{equation}\label{recursive}
\mathbf{P}^{s}=
\left[
  \begin{array}{cc}
    0 & \left(\bm{\alpha }^s\right)^T \\
    \bm{\alpha }^s & \mathbf{P}^{s-1} \\
  \end{array}
\right],\quad
\mathbf{b}^{s}=
\left[
  \begin{array}{cc}
    b_s & \mathbf{b}^{s-1} \\
  \end{array}
\right].
\end{equation}
The resultant length-$2^s$ RM sequence determined by
$\{\mathbf{P}^{s},\mathbf{b}^s\}$ according to
(\ref{RM_generate_recast}) is then termed as the \emph{$s$-th order
  sub-sequence} $\mathbf{c}^s$.

The nested structure of RM sequences is summarized in the following theorem.

\begin{figure}[!htp]
\small
\includegraphics[width=0.225\textwidth]{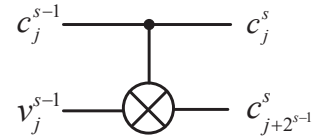}
\centering
\caption{The nested structure of RM sequences.}
\label{nested_structure}
\end{figure}

\noindent \textbf{Theorem 1:} As depicted in
Fig. \ref{nested_structure}, given the RM sequence $\mathbf{c}^m$, its
sub-sequences of order $s$ and $(s-1)$ satisfy the \textit{nested
  structure} that
\begin{equation}\label{theo_nested_structure}
\mathbf{c}^{s}=(\mathbf{c}^{s-1},\mathbf{c}^{s-1}\odot{\mathbf{v}^{s-1}}),2\leq{s}\leq{m}.
\end{equation}
The vector ${\mathbf{v}^{s-1}}$ is a length-$2^{s-1}$ Walsh sequence determined by the $s$-th layer of the matrix $\mathbf{P}^{m}$, and its $j$-th entry is given by
\begin{equation}\label{V_walsh}
v^{s-1}_j=(-1)^{b_s} \cdot (-1)^{(\bm{\alpha}^s)^T\mathbf{a}^{s-1}_{j-1}}=(-1)^{(\bm{\alpha}^s)^T\overline{\mathbf{a}^{s-1}_{j-1}}},
\end{equation}
where $\overline{\mathbf{a}^{s-1}_{j-1}}\triangleq{\mathbf{a}^{s-1}_{j-1}\oplus \mathbf{1}^{s-1}}$.

\begin{IEEEproof}
We prove that RM sequences obey the above nested structure relying on the generation
function given in (\ref{RM_generate_recast}).  When we have $1 \leq j \leq
2^{s-1}$, the vector $\mathbf{a}^s_{j-1}$ can be decomposed as
\begin{equation*}
\mathbf{a}^s_{j-1}=\left[
  \begin{array}{cc}
    0 & {\mathbf{a}^{s-1}_{j-1}} \\
  \end{array}
\right]^T.
\end{equation*}
Then the exponent of $c^{s}_j$ is expanded as
\begingroup
\allowdisplaybreaks
\begin{align*}
&(\mathbf{b}^s)^T\mathbf{a}^s_{j-1}+\frac{1}{2}(\mathbf{a}^s_{j-1})^T\mathbf{P}^s\mathbf{a}^s_{j-1}\\
=&\left[\begin{array}{cc}
b_s & (\mathbf{b}^{s-1})^T \\
\end{array}\right]
\left[ \begin{array}{c}
0 \\ \mathbf{a}^{s-1}_{j-1} \\
\end{array} \right]\\
&+\frac{1}{2} \left[\begin{array}{cc}
0 & (\mathbf{a}^{s-1}_{j-1})^T \\
\end{array}\right]
\left[\begin{array}{cc}
    0 & (\bm{\alpha}^s)^T \\
    \bm{\alpha}^s & \mathbf{P}^{s-1} \\
  \end{array}\right]
\left[ \begin{array}{c}
    0 \\ \mathbf{a}^{s-1}_{j-1} \\
  \end{array}\right]
\\
  =&(\mathbf{b}^{s-1})^T\mathbf{a}^{s-1}_{j-1}+\frac{1}{2}(\mathbf{a}^{s-1}_{j-1})^T\mathbf{P}^{s-1}\mathbf{a}^{s-1}_{j-1},
\end{align*}
\endgroup
hence we have $c^s_j=c^{s-1}_j$ in this case.
Similarly, by decomposing $\mathbf{a}^{s}_{j+2^{s-1}-1}$ as
\begin{equation*}
\mathbf{a}^{s}_{j+2^{s-1}-1}=\left[
  \begin{array}{cc}
    1 & {\mathbf{a}^{s-1}_{j-1}} \\
  \end{array}
\right]^T,
\end{equation*}
it is derived that
\begin{equation*}
\begin{aligned} &(\mathbf{b}^s)^T\mathbf{a}^s_{j-1}+\frac{1}{2}(\mathbf{a}^s_{j-1})^T\mathbf{P}^s\mathbf{a}^s_{j-1}\\
  =&(\mathbf{b}^{s-1})^T\mathbf{a}^{s-1}_{j-1}+\frac{1}{2}(\mathbf{a}^{s-1}_{j-1})^T\mathbf{P}^{s-1}\mathbf{a}^{s-1}_{j-1}
+b_s+(\bm{\alpha}^s)^T\mathbf{a}^{s-1}_{j-1}.
\end{aligned}
\end{equation*}
Therefore we have $c^{s}_{j+2^{s-1}}=c^{s-1}_j \cdot v^{s-1}_{j}$, where
\begin{equation*}
v^{s-1}_j=(-1)^{b_s} \cdot (-1)^{(\bm{\alpha}^s)^T\mathbf{a}^{s-1}_{j-1}}.
\end{equation*}
Combined with $b_s={(\bm{\alpha}^s)^T \mathbf{1}^{s-1}}$ from (\ref{b}), it is recast that $v^{s-1}_j=(-1)^{(\bm{\alpha}^s)^T\overline{\mathbf{a}^{s-1}_{j-1}}}$,
where $\overline{\mathbf{a}^{s-1}_{j-1}}\triangleq\mathbf{a}^{s-1}_{j-1}\oplus
\mathbf{1}^{s-1}$.

On this basis, we may conclude that
\begin{equation}\label{nested structure}
\left\{ \begin{aligned}
    & c^s_j=c^{s-1}_j \\
    & c^{s}_{j+2^{s-1}}=c^{s-1}_j \cdot v^{s-1}_{j}
  \end{aligned}\right.,1\leq j\leq2^{s-1},
\end{equation}
which completes the proof.
\end{IEEEproof}

\section{Blind Active User Detection and Channel Estimation Algorithm}
In this section, a novel RM detection algorithm is proposed for single
sequence detection.  Based on the nested structure of RM sequences, we recover the matrix-vector pair layer by layer and estimate the
channel coefficients in a straightforward manner.  Moreover, an enhanced
algorithm is conceived for handling error propagation.  In the case
where multiple RM sequences coexist, an iterative RM detection and
channel estimation algorithm is adopted.

\subsection{Single-Sequence Scenario: Layer-by-Layer RM Detection Algorithm}
\begin{figure}[!htp]
\small
\centering
\includegraphics[width=0.45\textwidth]{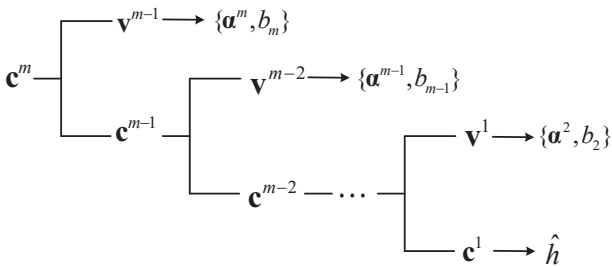}
\caption{The process of layer-by-layer RM detection algorithm.}
\label{algorithm process}
\end{figure}

When there is only a single active user in the system, the received signal can be rewritten as
\[ y^m_j=h\cdot c^m_j+e^m_j,j=1,\ldots,2^m.\]
The layer-by-layer RM detection algorithm is executed following the
process shown in Fig. \ref{algorithm process}.

Firstly, we split the
received signal into two length-$2^{m-1}$ parts and denote them as
${\left( {{{\mathbf{y}}^m}} \right)^\prime } = \left[ {y_1^m, \ldots
    ,y_{{2^{m - 1}}}^m} \right]^T$ and ${\left( {{{\mathbf{y}}^m}}
  \right)^{\prime \prime }} = \left[ {y_{{2^{m - 1}} + 1}^m, \ldots
    ,y_{{2^m}}^m} \right]^T$, respectively.  According to the nested
structure summarized in (\ref{nested structure}), the elements of
these two partial sequences are given by
\begin{equation*}
    \left\{ \begin{aligned}
  {\left( {y_j^m} \right)^\prime }&= hc_j^{m - 1} + {e^m_j} \hfill \\
  {\left( {y_j^m} \right)^{\prime \prime }}&= hc_j^{m - 1} \cdot v_j^{m - 1} + {e^m_{j + {2^{m - 1}}}} \hfill \\
\end{aligned}  \right.,j = 1, \ldots ,{2^{m - 1}}.
\end{equation*}
Then, we perform the element-wise conjugate multiplication on them, i.e.,
\begin{equation}\label{conj}
\tilde{y}^m_j={\left( {y_j^m} \right)^\prime }\cdot \left({\left( {y_j^m} \right)^{\prime \prime }}\right)^*=\left|h\right|^2 \cdot v^{m-1}_j+\tilde{\varepsilon}^m_j,
\end{equation}
where \[\tilde{\varepsilon}^m_j\triangleq hc^{m-1}_j(e^{m}_{j + {2^{m - 1}}})^* + {h^*}{c^{m-1}_j}{v^{m-1}_j}{e^m_j} + {e^m_j}(e^{m}_{j + {2^{m - 1}}})^*.\]
As shown in (\ref{V_walsh}), ${\mathbf{v}^{m-1}}$ is a Walsh sequence
with frequency $\bm{\alpha}^m$.  Hence, $\bm{\alpha}^m$ can be
recovered by performing the Walsh-Hadamard Transformation (\textbf{WHT})
on (\ref{conj}).  Specifically, the $m$-th order Hadamard matrix is
represented as $\mathbf{T}^m=\left[
  \mathbf{t}^m_{1},\ldots,\mathbf{t}^m_{i},\ldots,\mathbf{t}^m_{2^m}\right]^T$
and its elements are
$t^m_{i,j}=(-1)^{(\mathbf{a}^m_{i-1})^{T}\overline{\mathbf{a}^m_{j-1}}},i,j\in\{1,2,\ldots,2^m\}$.
Note that the Hadamard matrix we used here is obtained by flipping the
normalized one, whose element is
$(-1)^{(\mathbf{a}^m_{i-1})^{T}{\mathbf{a}^m_{j-1}}}$, up and down.
We denote the result of the transformation by the vector
$\mathbf{V}^{m-1}=\textbf{WHT}\left\{ \tilde{\mathbf{y}}^m
\right\}=\mathbf{T}^{m-1} \cdot \tilde{\mathbf{y}}^m$, whose $i$-th entry is
\begin{equation}\label{V}
\begin{aligned}
V^{m-1}_i=& \mathbf{t}^{m-1}_{i} \cdot \tilde{\mathbf{y}}^m \\
=& \left|h\right|^2 \sum\limits_{j = 1}^{{2^{m - 1}}} (-1)^{(\bm{\alpha}^m)^T\overline{\mathbf{a}^{m-1}_{j-1}}} \cdot (-1)^{(\mathbf{a}^{m-1}_{i-1})^{T}\overline{\mathbf{a}^{m-1}_{j-1}}}+\varepsilon^{m-1}_i\\
=& \left|h\right|^2 \sum\limits_{j = 1}^{{2^{m - 1}}} (-1)^{{(\bm{\alpha}^m+\mathbf{a}^{m-1}_{i-1})^T}\overline{\mathbf{a}^{m-1}_{j-1}}}+\varepsilon^{m-1}_i,
\end{aligned}
\end{equation}
where $\varepsilon^{m-1}_i\!\!\triangleq\!\!\sum\limits_{j = 1}^{{2^{m - 1}}}(-1)^{(\mathbf{a}^{m-1}_{i-1})^{T}\overline{\mathbf{a}^{m-1}_{j-1}}} \cdot \tilde{\varepsilon}^m_j $.
Thus, when we have $\mathbf{a}^{m-1}_{i-1}=\bm{\alpha}^m$, $V^{m-1}_i$ will achieve its maximum value, which equals to $2^{m-1}|h|^2+\varepsilon^{m-1}_i$.
On this basis, the vector $\bm{\alpha}^m$ can be recovered by searching through the vector $\mathbf{V}^{m-1}$ for the component having the largest value.
Additionally, for simplicity, we can opt for comparing the real parts of $\mathbf{V}^{m-1}$ with no performance loss.
Specifically, with
\begin{equation*}
w=\mathop {\arg \max }\limits_i \left\{\left(V^{m-1}_1\right)_I,\ldots,\left(V^{m-1}_{i}\right)_I,\ldots,\left(V^{m-1}_{2^{m-1}}\right)_I \right\},
\end{equation*}
 and $(\cdot)_I$ representing the real component, the estimate
$\hat{\bm{\alpha}}^m$ is the $(m-1)$-bit binary expression of $(w-1)$
and then $\hat{b}^m$ is the modulo-2 addition of  the entries in
$\hat{\bm{\alpha}}^m$, i.e.,
\begin{equation}\label{alpha_estimate}
\hat{\bm{\alpha}}^m=\mathbf{a}^{m-1}_{w-1},\quad\hat{b}_m=\hat{\alpha} _{1}^m \oplus \hat{\alpha } _{2}^m \oplus  \cdots  \oplus \hat{\alpha} _{m - 1}^m.
\end{equation}
 Upon substituting (\ref{alpha_estimate}) into (\ref{V_walsh}), the estimate $\hat{\mathbf{v}}^{m-1}$ is obtained.

Next, combining the estimate $\hat{\mathbf{v}}^{m-1}$ and  the pair of partial sequences $\left( {{{\mathbf{y}}^m}} \right)^\prime$ and $\left(
{{{\mathbf{y}}^m}} \right)^{\prime\prime}$, we  arrive at:
\begin{equation}\label{combine}
{{\mathbf{y}}^{m - 1}} = \frac{1}{2}\left( {{{\left( {{{\mathbf{y}}^m}} \right)}^\prime } + {{{\mathbf{\hat v}}}^{m - 1}} \odot {{\left( {{{\mathbf{y}}^m}} \right)}^{\prime \prime }}} \right).
\end{equation}
Under the condition that the estimate $\hat{\mathbf{v}}^{m-1}$ is correct, it is derived that
\begin{equation}\label{correct_combine}
y_j^{m - 1}= hc_j^{m - 1} + e^{m-1}_j,j=1,\ldots,2^{m-1},
\end{equation}
where $e^{m-1}_j \triangleq \dfrac{1}{2}\left( {{e^m_j} + \hat v_j^{m - 1} \cdot {e^m_{j + {2^{m - 1}}}}} \right)$.

In the sequel, the sequence $\mathbf{y}^{m-1}$ is split into
$\left(\mathbf{y}^{m-1}\right)^{\prime}$ and
$\left(\mathbf{y}^{m-1}\right)^{\prime\prime}$ and the estimates
$\left\{\hat{\bm{\alpha}}^{m-1},\hat{b}_{m-1}\right\}$ are calculated
by repeating the above steps.  This process continues until all the
estimates $\left\{
\hat{\bm{\alpha}}^{s},\hat{b}_s\right\},s\in\left\{2,\ldots,m\right\}$
and the length-2 sequence $\mathbf{y}^{1}\!\!=\!\!\left[ y^1_1,y^1_2
  \right]^T$ are obtained.

Then, the channel estimation can be done based on $\mathbf{y}^{1}$.
 Assuming that all the previous estimates are correct, we have
\begin{equation}
\left\{ \begin{gathered}
  y_1^1 =hc_1^1 + e^1_1=h+ e^1_1\hfill \\
  y_2^1 =hc_2^1 + e^1_2=h\cdot(-1)^{{b}_1}+ e^1_2 \hfill \\
\end{gathered}  \right.,
\end{equation}
where
\begin{equation*}
\left\{ \begin{gathered}
  e_1^1 =\frac{1}{{{2^{m - 1}}}}\left( {e_1^m \pm e_3^m \pm  \cdots  \pm e_{{2^m} - 1}^m} \right)\hfill \\
  e_2^1 =\frac{1}{{{2^{m - 1}}}}\left( {e_2^m \pm e_4^m \pm  \cdots  \pm e_{{2^m}}^m} \right) \hfill \\
\end{gathered}  \right.,
\end{equation*}
and the signs depend on the estimates $\hat{\mathbf{v}}^{s},2\leq
s\leq m-1$.  Since $e^m_j,j=1,\ldots,2^m$ are i.i.d. random variables
following the distribution $\mathcal{CN}(0,N_0)$, the channel coefficient can thus be estimated by
\begin{equation}\label{h_estimate}
\hat{h} = \frac{1}{2} \left( y^1_1+ (-1)^{\hat{b}_1} \cdot y^1_2 \right),
\end{equation}
where $\hat{b}_1=\hat{b}_m \oplus \cdots \oplus \hat{b}_2$ according to (\ref{b}).

Capitalizing on all the estimations above, the matrix-vector pair
$\left\{ \mathbf{P}^m,\mathbf{b}^m \right\}$ is recovered according to
(\ref{recursive}) and then reverts to the active user ID.  To
summarize, the layer-by-layer RM detection algorithm  conceived for single sequence detection is presented in \textbf{Algorithm 1}.

\begin{algorithm}[htb]
\caption{Layer-by-Layer RM Detection Algorithm}
\label{RMC_single}
\hspace*{0.02in}{\bf Input:}
the received signal $\mathbf{y}^m$;\quad\\
\hspace*{0.02in}{\bf Output:}
the matrix-vector pair $\{\hat{\mathbf{P}}^m,\hat{\mathbf{b}}^m\}$, the channel coefficient $\hat{h}$.
\begin{algorithmic}[1] 
\FOR{$s=m:-1:2$}
\STATE Split $\mathbf{y}^s$ into two partial sequences $\left( \mathbf{y}^{s} \right)^{\prime}$ and $\left( \mathbf{y}^{s} \right)^{\prime\prime}$.
\STATE Perform the element-wise conjugate multiplication on the $\left( \mathbf{y}^{s} \right)^{\prime}$ and $\left( \mathbf{y}^{s} \right)^{\prime\prime}$ with (\ref{conj}).
\STATE Perform the \textbf{WHT} and obtain the estimates $\{{\hat{\bm{\alpha}}^s},\hat{b}_s\}$ according to (\ref{alpha_estimate}).
\STATE Obtain ${{\mathbf{\hat v}}^{s - 1}}$ by applying $\{\hat{\bm{\alpha}}^s,\hat{b}_s\}$ to (\ref{V_walsh}).
\STATE Calculate the sequence $\mathbf{y}^{s-1}$ with (\ref{combine}).
\ENDFOR
\STATE With the element $\hat{b}_1=\hat{b}_m \oplus \cdots \oplus \hat{b}_2$, the channel coefficient is estimated according to (\ref{h_estimate}).
\STATE The matrix-vector pair $\{\hat{\mathbf{P}}^m,\hat{\mathbf{b}}^m\}$ is recovered based on all the estimates $\left\{ \hat{\bm{\alpha}}^{s},\hat{b}_s,\hat{b}_1\right\},s\in\left\{2,\ldots,m\right\}$ and then reverts to the user ID.
\RETURN $\{\hat{\mathbf{P}}^m,\hat{\mathbf{b}}^m\},\hat{h}$.
\end{algorithmic}
\end{algorithm}

 However, we found that Algorithm 1 suffers from an error propagation problem.
 To elaborate,  let us assume that an error occurs during the recovery of
$\left\{ {{{\bm{\alpha }}^s},{b_s}} \right\}$, which results in
${{\mathbf{\hat v}}^{s - 1}} \neq {{\mathbf{v}}^{s - 1}}$.  Then the
result of (\ref{combine}) turns out to be
\begin{equation}\label{error_occur}
\begin{aligned}
y_j^{s - 1}&= \frac{1}{2}\left( {{{\left( {y_j^s} \right)}^\prime } + \hat v_j^{s - 1} \cdot {{\left( {y_j^s} \right)}^{\prime \prime }}} \right) \\
&= \frac{1}{2}hc_j^{s - 1}\left( {1 + v_j^{s - 1}\hat v_j^{s - 1}} \right) + e^{s-1}_j\\
&=\left\{ \begin{aligned}
&hc_j^{s - 1} + {e^{s-1}_j},\quad&\text{if}\;v_j^{s - 1} = \hat v_j^{s - 1}\\
&{e^{s-1}_j},&\text{if}\;v_j^{s - 1} \ne \hat v_j^{s - 1}
\end{aligned} \right..
\end{aligned}
\end{equation}
Since ${{\mathbf{\hat v}}^{s - 1}}$ and ${{\mathbf{v}}^{s - 1}}$ are
two different Walsh sequences, the cardinality of the set
$J=\left\{j|v_j^{s - 1} \neq \hat v_j^{s - 1}\right\}$ definitely
equals to $2^{s-2}$.  As a result, the sequence $\mathbf{y}^{s-1}$
loses half  the information of $\mathbf{c}^{s-1}$, which will  impose consistent errors in the following steps.

In the sequel, we propose a list detection algorithm  for overcoming this error propagation problem.  The algorithm can be  partitioned into the extension and validation stages.

\begin{figure}[!htp]
\small
\centering
\includegraphics[width=0.42\textwidth]{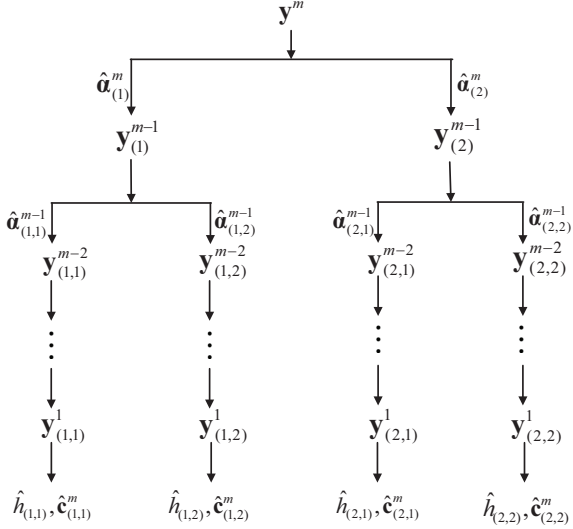}
\caption{The example of list detection algorithm with $\mathbf{L}=[2,2]$ and $F=2$.}
\label{list_detection}
\end{figure}

Firstly,  let us consider the extension stage.  For ease of  clarification, we
introduce two parameters $(\mathbf{L},F)$, where
$\mathbf{L}=\left[\ L_m,L_{m-1},\ldots,L_{m-F+1} \right]$ determines the width of extension and $F=\left| \mathbf{L} \right|$, i.e., the
number of elements in the vector $\mathbf{L}$, determines the depth of
extension.  On this basis, during the recovery of $\left\{
\bm{\alpha}^{s},b_s\right\}$ (line 4 of Algorithm 1), if $s\geq
m-F+1$, instead of choosing the location  having the largest real
component, we retain  the $L_s$ most likely locations to obtain a series of
candidates.  Then, the detection is continued in these parallel
threads, each called a detection path.  For the list detection
algorithm with parameters $(\mathbf{L},F)$, the total number of the
detection paths at the end equals to $\prod\nolimits_{s = m}^{m - F +
  1} {{L_s}}$, and the path  having the notation
$\mathbf{l}=(l_m,\ldots,l_{m-f+1},\ldots,l_{m-F+1})$ means that we
choose the $l_{m-f+1}$-th list in the $f$-th layer  along this path.
Intuitively, if $F=m-1$ and
$\mathbf{L}=\left[\ 2^{m-1},2^{m-2},\ldots,2 \right]$, the list
detection algorithm is equivalent to  the classic maximum likelihood sequence
estimation (MLSE), which  achieves the optimal performance with a
complexity that is super-linear to the size of  the sequence space.  Faced with the huge sequence space of RM sequences, the complexity of MLSE
is  excessive.  In the proposed list detection algorithm, we keep
track of several most likely paths to reach a compromise between
complexity and performance.  An example of $([2,2],2)$ list detection
algorithm is illustrated in Fig. \ref{list_detection}.

At the validation stage, the optimal path is chosen by measuring  the residual sequence energy.  To be specific, the residual sequence
energy on the path $\mathbf{l}$ is expressed as
\begin{equation}\label{residual_energy}
R_{\mathbf{l}}=\left\| \mathbf{y}^m-\hat{h}_{\mathbf{l}}\cdot \hat{\mathbf{c}}^m_{{\mathbf{l}}} \right\|^2.
\end{equation}
Then the path  having the minimum residual energy is decided to be the
final detection result.  The reasons to use  the residual energy as the
path metric are two fold.  On one hand, it is intuitive that if the
detected sequence  actually exists, the residual energy should
 be reduced after it is canceled from the received signal.  On the other
hand, if errors occur during the detection process, the channel
coefficient estimated by (\ref{h_estimate}) can be proved to be a very
small value.
 Specifically, if the algorithm is executed to the last step
based on the result derived in (\ref{error_occur}), the sequence $\mathbf{y}^1_1$ would have the form
\begin{equation*}
y_j^{1}=\left\{ \begin{array}{ll}
hc_j^{1} + {e^{1}_j},\quad &\text{if}\;v_j^{s^{\prime}} = \hat v_j^{s^{\prime}}, \forall s^{\prime}=s-1,\ldots,2\\
{e^{1}_j},&\text{otherwise}
\end{array} \right..
\end{equation*}
Since there is little chance that $v_j^{s^{\prime}} = \hat
v_j^{s^{\prime}}$ is satisfied for every $s^{\prime}\in
\{s-1,\ldots,2\}$ when errors occurred in the previous layers,  we have $y_j^{1} \approx e_j^{1} \approx 0$ and then $\hat{h}
\approx 0$ according to (\ref{h_estimate}).  Hence, the result
calculated by (\ref{residual_energy}) is almost equal to the energy of
the received signal and the corresponding path will be discarded.

\subsection{Multi-Sequence Scenario: Iterative RM Detection and Channel Estimation Algorithm}
In this section, we consider the scenario where multiple active users
coexist in the system.  We detect  the RM sequences and estimate  the channel coefficients in an iterative  manner. The process is described in detail below.

We first introduce some notations.
The maximum number of detected users is limited to $k_\text{max}$.
On one hand, a small value of $k_\text{max}$ would undoubtedly result in the unsatisfactory successful detection probability.
On the other hand, if $k_\text{max}$ is set to be too large, it may cause high false alarm rate (FAR) and computational complexity.
Hence, the choice of the exact value of $k_\text{max}$ is determined by the trade-off between the successful detection probability and the tolerable FAR, as well as the practical complexity constraint of the system.
For ease of exposition, the superscripts in $\mathbf{y}^m$ and $\mathbf{c}^m$ are dropped.
The stage where  the RM sequence of the detected user $k$ is updated during the $n$-th iteration is denoted as Stage $(n,k)$.
 Furthermore, $\mathbf{c}^{(n)}_k$ and $h^{(n)}_k$ represent the RM detection result and the channel estimate at Stage $(n,k)$, respectively.

Firstly,  we initialize the RM sequence and the channel coefficient as $h^{(0)}_k=0$ and $\mathbf{c}^{(0)}_k=\mathbf{0}^{2^m}$, respectively.
At Stage $(n,k)$, the multiple access interference (MAI) is expressed as
\begin{equation}\label{MAI}
\zeta_{k,j}^{(n)} = \sum\limits_ {k'\neq k} {h_{k'}^{(n')} \cdot c_{k',j}^{(n')}},
\end{equation}
where
\begin{equation*}
\left\{
\begin{array}{ll}
n'=n,\quad & \text{if}\;1\leq k' \leq k-1\\
n'=n-1, &\text{if}\;k+1\leq k' \leq  k_{\text{max}}
\end{array}
\right..
\end{equation*}
Then, the result of  $(y_j-\zeta_{k,j}^{(n)})$ is input to
\textbf{Algorithm 1} to obtain $\mathbf{c}^{(n)}_k$ and $h^{(n)}_k$.

Next, after all the  detected users' RM sequences are updated in this
iteration, the results are fed to the linear least square channel
estimator  for further improving their accuracy, which is specified as
\begin{equation}\label{LS}
\begin{aligned}
\mathbf{h}^{(n)}&=\left[ {h_1^{(n)}, \ldots ,h_{k_{\text{max}}}^{(n)}} \right]^T\\
&= \mathop {\arg \min }\limits_{\left[ {{h_1}, \ldots ,{h_{k_{\text{max}}}}} \right]^T \in {\mathbb{C}^{k_{\text{max}}}}} {\sum\limits_{j = 1}^{{2^m}} {\left\| {{y_j} - \sum\limits_{k' = 1}^{k_{\text{max}}} {{h_{k'}} \cdot c_{k',j}^{(n')}} } \right\|} ^2}.
\end{aligned}
\end{equation}
Thanks to the  information exchange between  the sequence detector and the channel estimator, the performance  gradually improves and the algorithm  is terminated when the results are converged or the maximum number of iteration is reached.
The multi-sequence detection  procedure is summarized in \textbf{Algorithm 2}.
\begin{algorithm}[htb]
\caption{ Iterative RM Detection and Channel Estimation Algorithm}
\label{RMC_multiple}
\hspace*{0.02in}{\bf Input:}\\
\hspace*{0.2in}the received signal $\mathbf{y}$,\\
\hspace*{0.2in}the maximum number of iteration $n_\text{max}$,\\
 \hspace*{0.2in}the maximum number of detected users $k_\text{max}$.\\
\hspace*{0.02in}{\bf Output:}\\
\hspace*{0.2in}the active user ID set $\mathbf{D}=\left\{D_1,\ldots,D_{k_{\text{max}}}\right\}$,\\
\hspace*{0.2in}the channel coefficient $\hat{\mathbf{h}}$.
\begin{algorithmic}[1] 
\STATE Initialization:\\[0.1in]
\hspace*{0.15in}$\mathbf{C}^{(0)}=\left[ \mathbf{c}^{(0)}_1,\ldots,\mathbf{c}^{(0)}_{k_{\text{max}}}\right]=\mathbf{0}^{2^m\times {k_{\text{max}}}}$, $\mathbf{h}^{(0)}=\mathbf{0}^{{k_{\text{max}}}}$.
\FOR {$n=1:n_{\text{max}}$}
\FOR {$k=1: k_{\text{max}}$}
\STATE Calculate the MAI $\zeta_{k,j}^{(n)}$ according to (\ref{MAI}).
\STATE Input the result of  $(y_j-\zeta_{k,j}^{(n)})$ to \textbf{Algorithm 1} and record the output as $\mathbf{c}^{(n)}_k$ and $h^{(n)}_k$.
\ENDFOR
\STATE Combine the RM detection results obtained in this iteration to the matrix $\mathbf{C}^{(n)}=\left[ \mathbf{c}^{(n)}_1,\ldots,\mathbf{c}^{(n)}_{k_{\text{max}}}\right]$ and feed it into the least square estimator specified in (\ref{LS}) to update all the active users' channel coefficients $\mathbf{h}^{(n)}$.
\ENDFOR
\STATE Obtain the active user ID set $\mathbf{D}$ based on $\mathbf{C}^{(n_{\text{max}})}$.
\RETURN $\mathbf{D}$, $\hat{\mathbf{h}}=\mathbf{h}^{(n_\text{max})}$.
\end{algorithmic}
\end{algorithm}

\section{Performance Analysis}

\subsection{Performance of Layer-by-Layer RM Detection Algorithm}
\subsubsection{Successful Detection Probability}
Without loss of generality, we assume that the transmitted RM sequence
is $\mathbf{c}^{m}=\mathbf{1}^{2^m}$ and the elements of the received
signal $y_j^m,j=1,\ldots,2^m$ are statistically independent.  To
derive the successful detection probability of single sequence
detection algorithm, we consider the  layer-by-layer recovery and the
following theorem concludes the correct recovery probability of each
layer.

\noindent \textbf{Theorem 2:}
Under the circumstance that we have recovered $\left\{ {{{\bm{{\alpha} }}^{m }},{{b}_{m }}} \right\}, \ldots, \left\{ {{{\bm{{\alpha} }}^{s+1}},{{b}_{s+1}}} \right\}$ correctly, the estimates $\left\{ {{{\bm{{\hat{\alpha}} }}^s},{{\hat{b}}_s}} \right\}$ are correct with the probability  of
\begin{equation}\label{Theorem2}
\small
{P^{(s|m, \ldots ,s + 1)}} \geq\max\left\{0,1 - ({2^{s - 1}} - 1)\Phi \left( {\frac{{0 - {2^{s - 1}}{{\left| h \right|}^2}}}{{\sqrt {{2^{s - 1}}{{\left( {\tilde{\sigma} _\varepsilon ^s} \right)}^2}} }}} \right)\right\},
\end{equation}
where ${\left( {\tilde{\sigma} _\varepsilon ^s} \right)^2} = 2{\left| h \right|^2}\dfrac{{{N_0}}}{{{2^{m - s}}}} + {\left( {\dfrac{{{N_0}}}{{{2^{m - s}}}}} \right)^2}$, $2\leq s\leq m$

and $\Phi(\cdot)$ is the distribution function of the standard normal distribution.
\begin{IEEEproof}
In the sequel, we prove \textbf{Theorem 2} by induction.
 Firstly, let us consider the case  of $s=m$.
Recall that $e^m_j,j=1,\ldots,2^m$ are i.i.d. random variables following $\mathcal{CN}(0,N_0)$ .
Given the channel coefficient $h$, the received signal obeys $y_j^m \sim \mathcal{CN}\left( {h,{N_0}} \right)$.
Then, upon approximating $\tilde{\varepsilon}^m_j$ in (9) by a complex Gaussian variable with mean $0$ and variance ${\left( {\tilde{\sigma} _{\varepsilon} ^m} \right)^2}\triangleq 2{\left| h \right|^2}{N_0}+ N_0^2$, we see that $\tilde{y}^m_j$ follows the distribution $\mathcal{CN}\left( {{{\left| h \right|}^2},{{\left( {\tilde{\sigma}_\varepsilon ^m} \right)}^2}} \right)$.

Next, the Walsh-Hadamard transformation is performed on
$\tilde{\mathbf{y}}^m$ and the result is shown in (\ref{V}).  When
$i=1$,  we have
\begin{equation}\label{V_m_1}
\small
V^{m-1}_1=2^{m-1}\cdot \left|h\right|^2+\sum\limits_{j = 1}^{{2^{m - 1}}} \tilde{\varepsilon}^m_j \sim \mathcal{CN}\left( {{2^{m - 1}}{{\left| h \right|}^2},{2^{m - 1}}{{\left( {\tilde{\sigma} _\varepsilon ^m} \right)}^2}} \right),
\end{equation}
while at other locations, i.e., $i=2,\ldots,2^{m-1}$,  we arrive at:
\begin{equation}\label{V_m_i}
V^{m-1}_i=\sum\limits_{j = 1}^{{2^{m - 1}}}(-1)^{(\mathbf{a}^{m-1}_{i-1})^{T}\overline{\mathbf{a}^{m-1}_{j-1}}} \cdot \tilde{\varepsilon}^m_j \sim \mathcal{CN}\left( {0,{2^{m - 1}}{{\left( {\tilde{\sigma} _\varepsilon ^m} \right)}^2}} \right).
\end{equation}
Since $\left\{ {{{\bm{\alpha }}^m},{b_m}} \right\}$ is recovered by searching for the specific element with the largest real component, once there exists some $i\in\{2,\ldots,2^{m-1}\}$ that results in the event ${A_{i}} =\left\{ \left( V^{m-1}_1 \right)_I-\left( V^{m-1}_i \right)_I<0\right\}$, the detection will head in the wrong direction.
On the basis of (\ref{V_m_1}) and (\ref{V_m_i}), the probability of the
event $A_{i}$ is
\begin{equation*}
\Pr ({A_{i}})=\Phi \left( {\frac{{0 - {2^{m - 1}}{{\left| h \right|}^2}}}{{\sqrt {{2^{m - 1}}{{\left( {\tilde{\sigma} _\varepsilon ^m} \right)}^2}} }}} \right),
\end{equation*}
and the estimates $\left\{ \hat{\bm{\alpha}}^m,\hat{b}_m \right\}$ are correct with the probability
\begin{equation}\label{P_correct}
{P^{(m)}}=1-{\bar P^{(m)}} =1- \Pr \left\{ {\bigcup\limits_{i= 2}^{{2^{m - 1}}} {{A_{i}}} } \right\}.
\end{equation}
 Using the classic union bound as the upper bound of the second term in (\ref{P_correct}) and considering that the probability $P^{(m)}$ cannot be less than 0, we have that
\begin{equation}\label{P_m}
\begin{aligned}
{P^{(m)}}&\geq \max\left\{ 0,1- \sum\limits_{i= 2}^{{2^{m - 1}}} {\Pr ({A_{i}})}\right\} \\
&= \max\left\{0,1 - ({2^{m - 1}} - 1)\Phi \left( {\frac{{0 - {2^{m - 1}}{{\left| h \right|}^2}}}{{\sqrt {{2^{m - 1}}{{\left( {\tilde{\sigma} _\varepsilon ^m} \right)}^2}} }}} \right)\right\},
\end{aligned}
\end{equation}
which is in accordance with (\ref{Theorem2}).

Next, assuming that ${P^{(s+1|m, \ldots ,s + 2)}}$ also agrees with
(\ref{Theorem2}) and the estimated $\left\{
\hat{\bm{\alpha}}^m,\hat{b}_m \right\},\ldots,\left\{
\hat{\bm{\alpha}}^{s+1},\hat{b}_{s+1} \right\}$ are correct,
${{{y}}^{s}_j}$ is calculated  as
\begin{equation}\label{correct_combine}
y_j^{s}= hc_j^{s} + e^{s}_j,j=1,\ldots,2^{s},
\end{equation}
where $e^{s}_j = \dfrac{1}{2}\left( {{e^{s+1}_j} + \hat v_j^{s} \cdot
  {e^{s+1}_{j + {2^{s}}}}} \right)$.  With the knowledge that
$e^{s+1}_j$ follows the distribution
$\mathcal{CN}\left(0,\dfrac{{{N_0}}}{2^{m-s-1}}\right)$,  we have that $y_j^{s}\sim
\mathcal{CN}\left(h,\dfrac{{{N_0}}}{2^{m-s}}\right)$.  The following
derivation is similar to the case  of $s=m$ and will not be repeated  here due to space limitations.
Finally, we can draw the conclusion that the correct  detection probability of $\left\{ \hat{\bm{\alpha}}^s,\hat{b}_s \right\}$ is identical  to (\ref{Theorem2}).
\end{IEEEproof}

The RM sequence $\mathbf{c}^m$ is detected successfully only if each
layer is correctly recovered, hence, its successful detection
probability is formulated as
\begin{equation*}
P = {P^{(m)}} \cdot {P^{(m - 1|m)}}\cdots {P^{(2|m,m - 1, \ldots ,3)}}.
\end{equation*}

As for the list detection algorithm, it is clear that its performance  is expected to be better than that of Algorithm 1, but inferior to that of MLSE.

\noindent\textit{Remark.}
\hspace*{0.08in}Taking a second look at the noise term in (\ref{correct_combine}) , we have:
\[e^{s}_j = \dfrac{1}{2}\left( {{e^{s+1}_j} + \hat v_j^{s} \cdot {e^{s+1}_{j + {2^{s}}}}} \right)\sim \mathcal{CN}\left(0,\dfrac{{{N_0}}}{2^{m-s}}\right),\]
 where we find that the noise level decreases at an exponential rate layer by layer.
Thus, if the first  few layers are recovered correctly, the subsequent recovery will be correct with a high probability.
This finding reveals the necessity to ensure the correctness of  the first few layers.  Inspired by this, to obtain  a better detection performance in the list detection algorithm, we can increase the number of lists  for the first few layers, while fixing the total number of paths.

\subsubsection{Computational Complexity}
{In the Layer-by-Layer RM  detection algorithm, the Walsh-Hadamard
transformation (WHT) plays the main role in the computational  tasks.
 Thus, we firstly consider the multiplication operations of  the WHT.
Thanks to the recursive structure of  the Walsh function,  a fast WHT can be utilized to reduce the complexity, which takes  $\mathcal{O}(s\cdot 2^s)$ multiplications when performed on a length-$2^s$ sequence.
Hence,  when only the WHT operations are considered, the total complexity of Algorithm 1 is $\mathcal{O}((m-2)2^m)$.
 By contrast, the algorithm given in \cite{generation_of_RM}  exhibits a complexity order up to $\mathcal{O} \left({{m^2} \cdot {2^m}} \right)$.
This is due to the fact that the algorithm in \cite{generation_of_RM} needs a length-$2^m$ fast WHT to recover each column of the matrix $\mathbf{P}^m$, while in Algorithm 1, the fast WHT is performed on $\tilde{\mathbf{y}}^s,s=m,m-1,\ldots,2$, whose length is decreasing exponentially layer by layer.
 Hence, our proposed Algorithm 1 has significant advantage  on complexity.
 If we further take the multiplications in Eq. (\ref{conj}) and Eq. (\ref{combine}) into account, the complexity of Algorithm 1 and the algorithm in \cite{generation_of_RM} is $\mathcal{O}\left[(m+1)\cdot 2^m\right]$ and $\mathcal{O} \left[(2m^2+m+2) \cdot {2^m}\right]$, respectively.

It is intuitive that the list detection algorithm improves the performance at the cost of an increased complexity. For the $(\mathbf{L},F)$ list detection algorithm, the number of  multiplications is given by
\begingroup
\allowdisplaybreaks
\begin{align*}
N_M(\mathbf{L},F)=&(m+2)2^{m-1}\\
&+\sum\limits^{m}_{s_1=m-F+1} \left( \prod\limits^{m}_{s_2=s_1} L_{s_2} \right){(s_1+1)2^{s_1-2}}\\
&+ \left(\prod\limits^{m}_{s_1=m-F+1}L_{s_1}\right) \sum\limits^{m-F}_{s_2=4}(s_2+1)2^{s_2-2}.
\end{align*}
\endgroup

 Zhang et al. \cite{RM_as_pilot} proposed a  shuffled version of the algorithm in \cite{generation_of_RM} , which aims to improve the performance by changing the recovery order of the columns of the matrix $\mathbf{P}^m$.  When the number of shuffles is $S$, the complexity of this enhanced algorithm equals to $\mathcal{O} \left[(2m^2+m+2) \cdot {2^m}S\right]$.}

Moreover, it is worth  to mention the  channel estimation complexity difference between these algorithms.
When executed  all the way to the last step,  Algorithm 1 can directly estimate the channel  coefficients according to (\ref{h_estimate}).
While the algorithms proposed in \cite{generation_of_RM} and \cite{RM_as_pilot} cannot compute the channel  coefficients and an extra least square (LS) estimation problem has to be solved, which incurs additional computational  complexity.

\subsection{Performance of  the Iterative RM Detection and Channel Estimation Algorithm}
For ease of  clarification, we analyze the performance in the case  of $K=2$.
At Stage $(1,1)$, the input of Algorithm 1 is exactly the received signal, i.e.,
\begin{equation}\label{received_signal_2}
y_j^{(1,1)} = {h_1} \cdot c_{1,j}^m + {h_2} \cdot c_{2,j}^m + {e^m_j},j = 1, \ldots ,{2^m}.
\end{equation}
After splitting it into  a pair of partial sequences and computing their element-wise conjugate multiplication,  we have
\begin{equation}\label{conj_2}
\begin{aligned}
\tilde{y}^m_j=&{\left| {{h_1}} \right|^2} \cdot v_{1,j}^{m - 1} + {\left| {{h_2}} \right|^2} \cdot v_{2,j}^{m - 1} + h_1^*{h_2}c_{1,j}^{m - 1}c_{2,j}^{m - 1}v_{1,j}^{m - 1} \\
&+ {h_1}h_2^*c_{1,j}^{m - 1}c_{2,j}^{m - 1}v_{2,j}^{m - 1} + {\tilde{\varepsilon}^m_j},j = 1, \ldots ,{2^{m - 1}},
\end{aligned}
\end{equation}
where ${\tilde{\varepsilon}^m_j}$ follows the distribution $\mathcal{CN}\left( {0,{\left( {\tilde{\sigma} _\varepsilon ^m} \right)^2}} \right)$ and ${\left( {\tilde{\sigma }_\varepsilon ^m} \right)^2}\!\!\triangleq\!\!2\left( {{{\left| {{h_1}} \right|}^2} + {{\left| {{h_2}} \right|}^2}} \right){N_0} + N_0^2$.
Then, the Walsh-Hadamard transformation is  applied to (\ref{conj_2}) for obtaining the vector $\mathbf{V}^{m-1}$.
Under the assumption that $\left| {{h_1}} \right|^2\geq \left| {{h_2}} \right|^2$, ${\bm{\alpha }}_1^m$ is recovered correctly if and only if ${\left( V^{m-1}_{\langle {\bm{\alpha }}_1^m \rangle} \right)_I}> {\left( {V_i^{m - 1}} \right)_I}$ is satisfied for every $i\in\{1,\ldots,2^{m-1}\}\backslash{\langle {\bm{\alpha }}_1^m \rangle}$, where ${\langle {\bm{\alpha }}_1^m \rangle}$ is the location index that satisfies ${\bm{\alpha }}_1^m=\mathbf{a}^{m-1}_{\langle {\bm{\alpha }}_1^m \rangle-1}$.

 Firstly, in the special case  of $\bm{\alpha }_1^m = \bm{\alpha}_2^m $, the real component of the element at the location $\langle {\bm{\alpha }}_1^m \rangle$ turns out to be
\begin{equation}\label{b1_b2}
\begin{aligned}
&{\left( V^{m-1}_{\langle {\bm{\alpha }}_1^m \rangle} \right)_I} \\
&= {2^{m - 1}}\left[(-1)^{b_{1,m}}\cdot{\left| {{h_1}} \right|^2+(-1)^{b_{2,m}}\cdot\left| {{h_2}} \right|^2} \right]\\
&+ \left[(-1)^{b_{1,m}}+(-1)^{b_{2,m}}\right]{\left( {h_1^*{h_2}} \right)_I}\sum\limits_{j = 1}^{{2^{m - 1}}}\left( {c_{1,j}^{m - 1}c_{2,j}^{m - 1}}\right)\\
&+ {\left( (\varepsilon^m_{1,j})^{\prime} \right)_I}.
\end{aligned}
\end{equation}
The benefit of our proposed mapping rule manifests itself here.
It is obvious from (\ref{b1_b2}) that if $b_{1,m}\neq b_{2,m}$, the component $\left( V^{m-1}_{\langle {\bm{\alpha }}_1^m \rangle} \right)_I$ will decrease to ${2^{m - 1}}\left({\left| {{h_1}} \right|^2-\left| {{h_2}} \right|^2} \right)+ {\left( (\varepsilon^m_{1,j})^{\prime} \right)_I}$ and there is a substantial chance that it will become smaller than the other interference components, which results in the incorrect recovery of $\bm{\alpha}^m_1$.
While our constraint in (\ref{b}) ensures that $b_{1,m}= b_{2,m}$ in this case and the magnitude of $\left( V^{m-1}_{\langle {\bm{\alpha }}_1^m \rangle} \right)_I$ is reinforced to be
\begingroup\label{V_reinforce}
\allowdisplaybreaks
\begin{align}\notag
{\left( V^{m-1}_{\langle {\bm{\alpha }}_1^m \rangle} \right)_I} &= {2^{m - 1}}\left({\left| {{h_1}} \right|^2+\left| {{h_2}} \right|^2} \right)\\
&+ 2{\left( {h_1^*{h_2}} \right)_I}\sum\limits_{j = 1}^{{2^{m - 1}}}\left( {c_{1,j}^{m - 1}c_{2,j}^{m - 1}}\right)
+ {\left( (\varepsilon^m_{1,j})^{\prime} \right)_I},
\end{align}
\endgroup
which is expected to improve the detection performance.

Next, in the case  of $\bm{\alpha }_1^m \neq \bm{\alpha }_2^m $, the real component of the element $V^{m-1}_{\langle {\bm{\alpha }}_1^m \rangle}$ is specified as
\begin{equation}\label{V1}
\begin{aligned}
&{\left( V^{m-1}_{\langle {\bm{\alpha }}_1^m \rangle} \right)_I} = {2^{m - 1}}{\left| {{h_1}} \right|^2}\\
&+ {\left( {h_1^*{h_2}} \right)_I}\sum\limits_{j = 1}^{{2^{m - 1}}}\Big({c_{1,j}^{m - 1}c_{2,j}^{m - 1}}+{c_{1,j}^{m - 1}v_{1,j}^{m - 1}} \cdot  c_{2,j}^{m - 1}v_{2,j}^{m - 1}\Big)\\
&+ {\left( (\varepsilon^m_{1,j})^{\prime} \right)_I},
\end{aligned}
\end{equation}
where $(\varepsilon^m_{1,j})^{\prime}\triangleq\sum\limits_{j = 1}^{{2^{m - 1}}} {{\tilde{\varepsilon}^m_j} \cdot v_{1,j}^{m - 1}} \sim \mathcal{CN}\left( {0,{2^{m - 1}}{{\left( {\tilde{\sigma} _\varepsilon ^m} \right)}^2}} \right)$.
The second term of (\ref{V1}) happens to be the inner product of $\mathbf{c}^{m}_1$ and $\mathbf{c}^{m}_2$, which is denoted as $\chi^{m}_{1,2}$.
Then, Eq. (\ref{V1}) can be recast as
\begin{equation}\label{V1_rewritten}
{\left( V^{m-1}_{\langle {\bm{\alpha }}_1^m \rangle} \right)_I} = {2^{m - 1}}{\left| {{h_1}} \right|^2} + {\left( {h_1^*{h_2}} \right)_I}\cdot \chi^{m}_{1,2} + {\left( (\varepsilon^m_{1,j})^{\prime} \right)_I}.
\end{equation}
 Similarly, we have that
\begin{equation*}
{\left( V^{m-1}_{\langle {\bm{\alpha }}_2^m \rangle} \right)_I} = {2^{m - 1}}{\left| {{h_2}} \right|^2} + {\left( {h_1^*{h_2}} \right)_I}\cdot \chi^{m}_{1,2} + {\left( (\varepsilon^m_{2,j})^{\prime} \right)_I}.
\end{equation*}
 While at the location $i\;\left(i\neq {\langle {\bm{\alpha }}_1^m \rangle}\:\text{and}\: i\neq{\langle {\bm{\alpha }}_2^m \rangle}\right)$, the real component is expressed as
\begin{equation*}
\small
\begin{aligned}
    &{\left( {V_i^{m - 1}} \right)_I} ={\left( {h_1^*{h_2}} \right)_I}\\
    &\cdot\sum\limits_{j = 1}^{{2^{m - 1}}}\Big(c_{1,j}^{m - 1}c_{2,j}^{m - 1} \cdot v_{1,j}^{m - 1} \cdot t_{i,j}^{m - 1}+ c_{1,j}^{m - 1}c_{2,j}^{m - 1}\cdot v_{2,j}^{m - 1} \cdot t_{i,j}^{m - 1}\Big)\\
    &+ {\left( (\varepsilon^m_{i,j})^{\prime} \right)_I},
\end{aligned}
\end{equation*}
recalling that $t_{i,j}^{m - 1}$ is the element of the Hadamard matrix $\mathbf{T}^{m-1}$.
To proceed, we recast the first term as follows:
\begin{align*}
&\sum\limits_{j = 1}^{{2^{m - 1}}}\left(c_{1,j}^{m - 1}c_{2,j}^{m - 1}v_{1,j}^{m - 1} t_{i,j}^{m - 1}+ c_{1,j}^{m - 1}c_{2,j}^{m - 1}v_{2,j}^{m - 1}t_{i,j}^{m - 1}\right)\\
=&\sum\limits_{j = 1}^{{2^{m - 1}}}\left(c_{1,j}^{m - 1}c_{2,j}^{m - 1}+c_{1,j}^{m - 1}v_{1,j}^{m - 1}c_{2,j}^{m - 1}t_{i,j}^{m - 1}\right)\\
&-\sum\limits_{j = 1}^{{2^{m - 1}}}\left(c_{1,j}^{m - 1}c_{2,j}^{m - 1}\right)\\
&+\sum\limits_{j = 1}^{{2^{m - 1}}} \left(c_{1,j}^{m - 1}c_{2,j}^{m - 1}+ c_{1,j}^{m - 1}t_{i,j}^{m - 1}c_{2,j}^{m - 1}v_{2,j}^{m - 1}\right)\\
&-\sum\limits_{j = 1}^{{2^{m - 1}}} \left(c_{1,j}^{m - 1}c_{2,j}^{m - 1}\right)\\
=&\:\chi^{m}_{1,(2,i)}-\chi^{m-1}_{1,2}+\chi^{m}_{(1,i),2}-\chi^{m-1}_{1,2},
\end{align*}
where $\chi^{m}_{1,(2,i)}$ denotes the inner product of $\mathbf{c}^{m}_{1}$ and the RM sequence generated by the matrix-vector expressed as
\begingroup
\allowdisplaybreaks[4]
\begin{align}\notag
\mathbf{P}^{m}_{(2,i)}=
&\left[
  \begin{array}{cc}
    0 & \left(\mathbf{a}^{m-1}_{i-1}\right)^T \\
    \mathbf{a}^{m-1}_{i-1} & \mathbf{P}^{m-1}_{2} \\
  \end{array}
\right],\\
\mathbf{b}^{m}_{(2,i)}=&\left[{a}^{m-1}_{i-1,1} \oplus \cdots \oplus {a}^{m-1}_{i-1,m-1},  \mathbf{b}^{m-1}_{2} \right],\notag
\end{align}
\endgroup
and $\chi^{m}_{(1,i),2}$ is obtained in the similar way.

It is obvious that the inner products of  the RM sequences play an
important role in  determining the performance of multi-sequence detection.
The  authors of \cite{generation_of_RM} reveal that the distribution of the inner products of second-order RM sequences is symmetric.
Moreover, if two distinct matrices $\mathbf{P}^{m}_{1}$ and $\mathbf{P}^{m}_{2}$
satisfy that
\begin{equation*}
    \text{rank}(\mathbf{P}^{m}_{1}-\mathbf{P}^{m}_{2})=2r, r=1,\ldots,[m/2],
\end{equation*}
then the magnitude of the inner product of their respective generated RM sequences is either $2^{m-r}$ or $0$.
 This inspires us to  restrict the matrices, for example, to be taken from the $DG(m,2r)$ set \cite{DG},  for further improving the  attainable performance of RM detection.

Then, on the premise that $\left\{\bm{\alpha}^{m}_1,b_{m,1}\right\}$ is correctly recovered, the sequence $\mathbf{y}^{m-1}$ is given as
\begin{equation*}
y_j^{m - 1} = {h_1} \cdot c_{1,j}^{m - 1} + \frac{1}{2}{h_2}c_{2,j}^{m - 1}\left( {1 + v_{1,j}^{m - 1} \cdot v_{2,j}^{m - 1}} \right) + e_j^{m - 1},
\end{equation*}
where $e_j^{m - 1} \!\!=\!\!\dfrac{1}{2}\left( {e_j^m + v_{1,j}^{m - 1} \cdot e_{j + {2^{m - 1}}}^m} \right)$.
Combined with (\ref{received_signal_2}), it is found that for the first active user, the signal-to-interference-and-noise-ratio (SINR) is enhanced.
Furthermore, as the algorithm proceeds, the SINR of $\mathbf{y}^{s}$, which is denoted as $\gamma ^s$, doubles layer by layer, specifically,
\begin{equation}\label{SINR}
{\gamma ^s} = \frac{2^{m-s}{{{\left| {{h_1}} \right|}^2}}}{{{{\left| {{h_2}} \right|}^2} + {N_0}}},s=m,\ldots,1.
\end{equation}
This finding further  underlines the necessity  of ensuring the correctness of the first few layers.

At the end of Stage $(1,1)$, the estimates $\mathbf{c}^{(1)}_1$ and $h^{(1)}_1$ are obtained.
Then we enter  Stage $(1,2)$.
 At the beginning of Stage $(1,2)$, the residual signal is calculated by
\begin{equation}\label{y_1_2}
y^{(1,2)}_{j}={h_2}c_{2,j}^m +\left({h_1}c_{1,j}^m -h^{(1)}_1 c^{(1)}_{1,j}\right) + {e^m_j},j = 1, \ldots ,{2^m},
\end{equation}
which is then input to  Algorithm 1.
The accuracy of  both $\mathbf{c}^{(1)}_1$ and $h^{(1)}_1$ affects the performance at this stage.
If the sequence $\mathbf{c}^{(1)}_1$ is correct, the
impact of the channel  estimation error (CEE) is relatively small.  On
one hand, in the case that $\bm{\alpha}^{s}_{1},s=m,\ldots,2$ are all
correctly recovered, the SINR of the last layer is up to
$\frac{2^{m-1}{{{\left| {{h_1}} \right|}^2}}}{{{{\left| {{h_2}}
        \right|}^2} + {N_0}}}$ according to (\ref{SINR}), which
ensures that the CEE will not be too  high.  On the other hand, the LS
channel estimator at the end of the first iteration  is capable of further improving the accuracy of $h^{(1)}_1$, and the second user's RM sequence
can be updated in the second iteration.

Whereas if the sequence $\mathbf{c}^{(1)}_1$ is incorrect,
(\ref{y_1_2}) is equivalent to  adding an extra sequence to the original
received signal, which further lower the SINR.  In
\cite{generation_of_RM} and \cite{RM_as_pilot},  a traditional SIC-based
method is used for multi-sequence detection .
 Explicitly, before detecting the next sequence, all the sequences obtained so far are subtracted from the received signal.
In this scheme, the impact of  any incorrectly detected sequences always persists.
 By contrast, the  iterative method adopted in our proposed algorithm deals with this problem in a better way.
For example, if the detected sequence $\mathbf{c}^{(1)}_1$ is wrong
but $\mathbf{c}^{(1)}_2$ happens to be right, then at Stage $(2,1)$,
after subtracting the MAI term, the signal input to Algorithm 1 is
equal to
\begin{equation*}
{h_1}c_{1,j}^m +\left({h_2}c_{2,j}^m -h^{(1)}_2 c^{(1)}_{2,j}\right) + {e^m_j},j = 1, \ldots ,{2^m},
\end{equation*}
which is free from the impact of the incorrect result $\mathbf{c}^{(1)}_1$ and $h^{(1)}_1$.
Moreover, the  algorithms in \cite{generation_of_RM} and \cite{RM_as_pilot} need to perform LS channel estimation after each sequence detection, while our proposed algorithm executes LS channel estimation only once in an iteration, which further reduces the complexity.

 Finally, the properties of our proposed algorithms are summarized as follows:
\begin{itemize}

  \item  As a benefit of the proposed mapping method, the channel coefficients can be estimated directly according to (\ref{h_estimate}), which leads to a reduced computational complexity without  any performance loss.
  \item In the case where multiple active users possess the same $\bm{\alpha}^s_k$, our proposed mapping rule  ensures that the corresponding result of the WHT is reinforced like in (\ref{V_reinforce}). As such, the successful recovery probability of $\bm{\alpha}^s_k$ is expected to be enhanced.
  \item  By exploiting the nested structure of RM sequences, the matrix-vector pair is recovered layer by layer. During this process, the interference and noise level is found to keep  on decreasing on the premise that $\left\{\bm{\alpha}^{s},b_{s}\right\}$ are recovered correctly. On one hand, this ensures the accuracy of channel estimation based on (\ref{h_estimate}). On the other hand,  this reminds us  of the importance of improving the recovery reliability of the first  few layers. Inspired by this, we can adjust the parameters of  the list detection algorithm  for enhancing the performance.
  \item The inner products of RM sequences have  a significant influence on the performance of multi-sequence detection. Motivated by this, we can further restrict the choice of the matrices $\mathbf{P}^m$ to enhance detection capability. Intuitively,  this is achieved at the cost of shrinking the user space size.
   \item   In contrast to the common SIC-based detection, the iterative detection algorithm utilized in this paper  is capable of reducing the impact of incorrectly detected sequences, thus it has  a better capability to deal with  the error propagation problem  of SIC.
\end{itemize}

\section{Simulation Results}
In this section, we evaluate the performance of  the proposed algorithms in terms of their sequence detection accuracy, channel estimation accuracy and computational cost.
For ease of exposition, Algorithm 1 and its list detection version are denoted as ``\emph{RM\_LLD}'' and ``\emph{list  RM\_LLD}'', respectively.
Moreover, the algorithm in \cite{generation_of_RM} is used as the benchmark.
Since it is based on the application of shift-and-multiply to the received signal, we referred to it and its  shuffled version proposed in \cite{RM_as_pilot} as ``\emph{RM\_SMD}'' and ``\emph{RM\_SMD with shuffle}'', respectively.
In the sequel, the successful detection probability is defined as the average ratio between the number of successfully detected active users and the  total number of active users in the system.
 The channel estimation error is  quantified by the mean square error of channel estimates of successfully detected active users.

\subsection{Performance of  the Layer-by-Layer RM Detection Algorithm}

Firstly, we examine the performance of the proposed algorithms in the single-sequence scenario. The simulation parameters are listed in Table. \ref{sp1}.
\begin{table}[htb]
\centering
\caption{Simulation Parameters of Single-Sequence Detection.}
\begin{tabular}{!{\vrule width0.8pt}l!{\vrule width0.8pt}l!{\vrule width0.8pt}}
  \Xhline{0.8pt}
  {Sequence Length} & {$2^8$} \\
  \hline
  {SNR} & $-10\sim 5$dB  \\
  \hline
  {Shuffle Times in ``\emph{RM\_SMD with shuffle}}'' & 4\\
  \hline
  {$(\mathbf{L},F)$ of ``\textit{list RM\_LLD}}'' & $([2,2],2)$\\
  \hline
  {Simulation Times} & $2000$ \\
  \Xhline{0.8pt}
\end{tabular}
\label{sp1}
\end{table}

\begin{figure}[!htp]
\small
\centering
\includegraphics[width=0.48\textwidth]{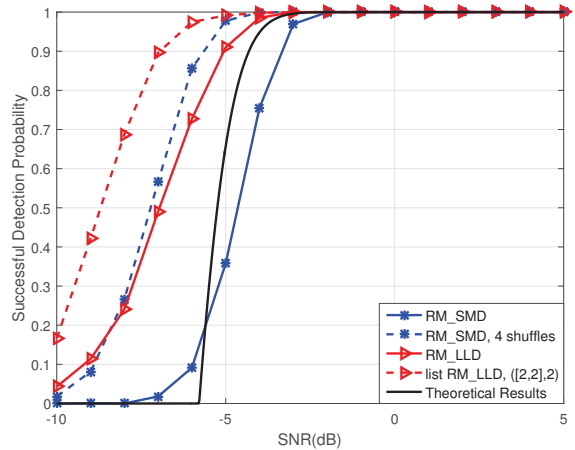}
\caption{The successful reconstruction probability versus the
signal to noise ratio (SNR) when $m=8$.}
\label{P_suc_SNR}
\end{figure}

 Fig. \ref{P_suc_SNR} illustrates the successful detection probability versus the signal to noise ratio (SNR).
It can be observed that \emph{RM\_SMD} faces severe performance degradation in the low-SNR scenario, while \emph{RM\_LLD}  exhibits robustness against the SNR, thus confirming the superior detection performance of the proposed algorithm.
As expected,  both the shuffling operation and the list detection approach are capable of improving the detection performance.
 Even though both \textit{RM\_SMD with 4 shuffles} and \textit{list RM\_LLD  with $([2,2],2)$} choose the optimal result from 4 candidates, it is shown  in Fig. \ref{P_suc_SNR} that \textit{list RM\_LLD with  $([2,2],2)$} is superior.
Moreover, the theoretical  results obtained in Section \uppercase\expandafter{\romannumeral5}.A is also depicted in Fig. \ref{P_suc_SNR}.
 Since the union bound  is utilized in the analysis, this result can act as the lower bound.
The simulation result of \emph{RM\_LLD} and the theoretical result consistently show that when  the SNR is higher than $-4$dB, the RM sequence can be  successfully detected with a probability close to $1$.

The computational complexity of these four algorithms  associated with $m=8$ is compared in  Table \ref{computational_complexity}.
Here we  quantify the complexity by the number of multiplication operations.
 Observe that the proposed \emph{RM\_LLD}  impose a much lower complexity than \emph{RM\_SMD}.
It is intuitive that  both the list detection approach and  the shuffle operation improve the performance  at the cost of an increased complexity.
However, the complexity of \emph{RM\_SMD with shuffles} increases linearly with the number of shuffles, while the complexity growth rate of \emph{list RM\_LLD} is much more gradual.
\begin{table}[htb]
\centering
\caption{Comparison of different algorithms in terms of computational complexity in the case of $m=8$.}
\begin{tabular}{!{\vrule width0.8pt}l!{\vrule width0.8pt}c!{\vrule width0.8pt}}
  \Xhline{0.8pt}
  Algorithm & Computational Complexity \\
  \Xhline{0.8pt}
  {\emph{RM\_LLD}} & 2304 \\
  \hline
  {\emph{list RM\_LLD, $([2,2],2)$}} & 3820  \\
  \hline
  {\emph{RM\_SMD}} & 35328\\
  \hline
  {\emph{RM\_SMD, 4 shuffles}} & 141312 \\
  \Xhline{0.8pt}
\end{tabular}
\label{computational_complexity}
\end{table}

\begin{figure}[!htp]
\small
\centering
\includegraphics[width=0.48\textwidth]{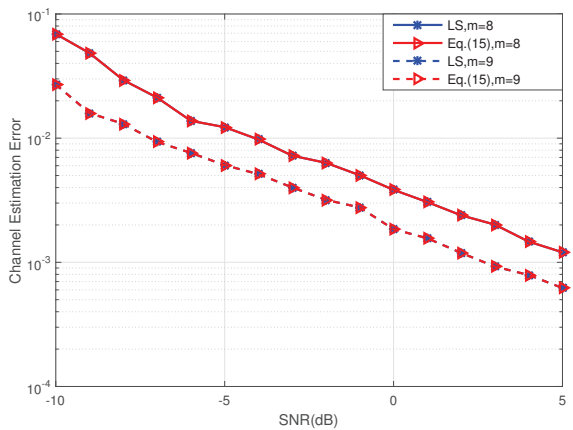}
\caption{The channel coefficient estimation error versus SNR.}
\label{CE2_SNR}
\end{figure}

 Finally, we verify the feasibility of estimating the channel  coefficients according to (\ref{h_estimate}).
 Fig. \ref{CE2_SNR} depicts the channel estimation errors  calculated both from Eq. (\ref{h_estimate}) and  by the least square (LS) estimator.
It can be seen from  Fig. \ref{CE2_SNR} that the channel estimates calculated by (\ref{h_estimate})  exhibit the same accuracy as the results obtained by the LS estimator.
However, it is obvious that the complexity of (\ref{h_estimate}) is much lower, which further validates the superiority of our proposed algorithm.

\subsection{Performance of  the iterative RM Detection and Channel Estimation Algorithm}
 Next, we evaluate the performance of our algorithms in the multi-sequence scenario and the corresponding parameters are summarized in Table. \ref{sp2}.
\begin{table}[htb]
\centering
\setlength\extrarowheight{2pt}
\caption{ Simulation Parameters of Multi-Sequence Detection.}
\begin{tabular}{!{\vrule width0.8pt}l!{\vrule width0.8pt}l!{\vrule width0.8pt}}
    \Xhline{0.8pt}
  {Sequence Length} & {$2^8$, $2^{10}$} \\
  \hline
  {SNR} & $20$dB  \\
  \hline
  \tabincell{l}{\# of iterations in \\``\emph{RM\_LLD, iterative}''} & 5  \\
  \hline
  {False Alarm Rate} & $\leq 10^{-7}$  \\
  \hline
  \tabincell{l}{Shuffle Times in \\``\emph{RM\_SMD with shuffle}''} & \tabincell{l}{4 for length-$2^8$ sequences;\\8 for length-$2^{10}$ sequences}\\
  \hline
  \tabincell{l}{$(\mathbf{L},F)$ of ``\textit{list RM\_LLD}''} & \tabincell{l}{$([2,2],2)$ and $([4],1)$ \\for length-$2^8$ sequences;\\$([4,2],2)$ and $([8],1)$ \\for length-$2^{10}$ sequences}\\
  \hline
  {Simulation Times} & $2000$ \\
   \Xhline{0.8pt}
\end{tabular}
\label{sp2}
\end{table}
\begin{figure}
\centering
\subfigure[$m=8$]{\includegraphics[width=0.4\textwidth]{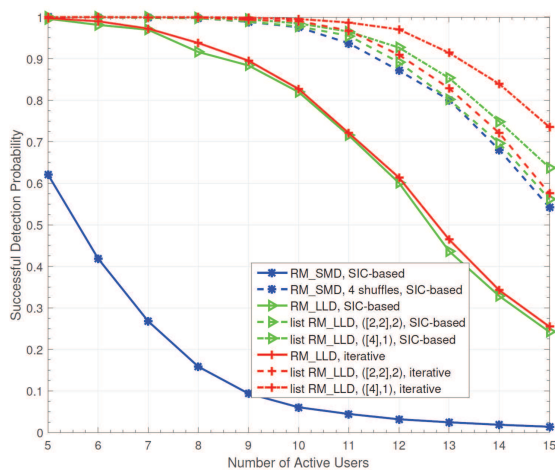}}
\subfigure[$m=10$]{\includegraphics[width=0.4\textwidth]{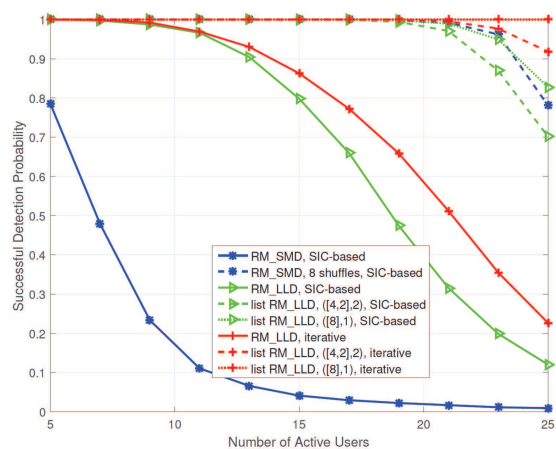}}
\caption{The successful detection probability versus the number of active users in the system.}
\label{P_suc_K}
\end{figure}


The successful detection probability versus the number of active users in the case of $m=8$ and $m=10$  is depicted in  Fig. \ref{P_suc_K}.
It can be observed that the \textit{RM\_SMD} curve drops rapidly  when increasing the number of active users, while \textit{RM\_LLD} shows  a significant advantage over it.
As in the single sequence scenario,  both ``\textit{RM\_SMD with shuffle}'' and ``\textit{list RM\_LLD}'' improve the successful detection probability.
Comparing the successful detection probability of ``\textit{list RM\_LLD, $([2,2],2)$}'' and ``\textit{list RM\_LLD, $([4],1)$}'', we find that with the total number of paths fixed, the performance can be  beneficially improved by increasing the number of lists in the first layer.
This result further validates the importance  of ensuring the reliable recovery of the first few layers in the proposed RM detection algorithm.
Moreover, it can be seen from Fig. \ref{P_suc_K} that the iterative method adopted by our proposed algorithm is superior  to the traditional SIC-based multi-sequence detection.
 This performance improvement is an explicit benefit of the mutual information exchange between the RM detector and the channel estimator as well as of the capability to eliminate the impact of incorrectly detected sequences.
Combining Fig. \ref{P_suc_K}(a) and (b), it is  also clear that the detection capability is enhanced by increasing the length of  sequences.
\begin{figure}[!htp]
\small
\centering
\includegraphics[width=0.48\textwidth]{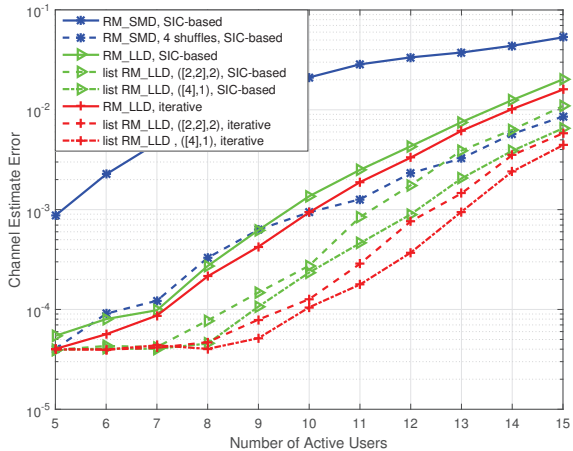}
\caption{The channel coefficient estimation error versus the number of active users in the system in the case of $m=8$.}
\label{CE2_ave_m8}
\end{figure}
Fig. \ref{CE2_ave_m8} compares the channel estimate error of different algorithms in the multi-sequence scenario when $m=8$.
An insightful observation is that although the successful detection probability of ``\textit{list RM\_LLD, SIC-based}'' and ``\textit{list RM\_LLD, iterative}'' seems similar in Fig. \ref{P_suc_K} (a), the channel estimates of ``\textit{list RM\_LLD, iterative}''  are obviously more accurate than  those of ``\textit{list RM\_LLD, SIC-based}''.
 Furthermore, the advantage of  the iterative method becomes more prominent for ``\textit{list RM\_LLD, $([2,2],2)$}'' and ``\textit{list RM\_LLD, $([4],1)$}''.
Based on these results, we may conclude that the proposed iterative RM detection and channel estimation algorithm  beneficially improves the user detection and channel estimation performance.

\begin{figure}[!htp]
\small
\centering
\includegraphics[width=0.48\textwidth]{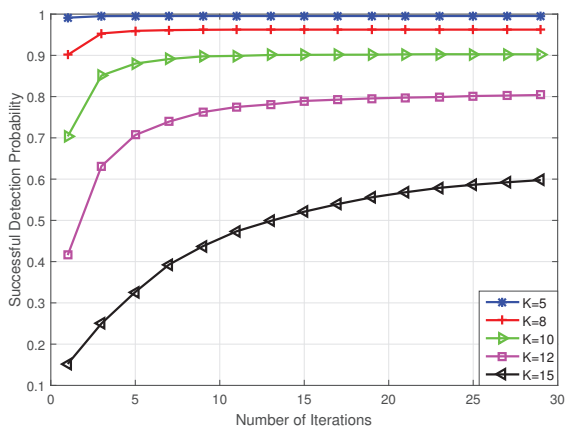}
\caption{ The convergence of the iterative RM detection and channel estimation algorithm.}
\label{convergence}
\end{figure}

Fig. \ref{convergence} depicts the convergence of the iterative RM detection and channel estimation algorithm for $m=8$.
It can be seen that the convergence rate is related to the number of active users in the system.
In the case of $K\leq 10$, the algorithm can converge within $5$ iterations.
Although it needs more iterations to converge when the number of active users increases, the successful detection probability keeps rising before achieving the convergence.
Hence, the value of $n_\text{max}$ can be set according to the practical tradeoff between the detection capability and the complexity.

\section{Conclusions}
In this paper, we utilize RM sequences for user identification and channel estimation to meet the massive connectivity and low latency requirements  of mMTC scenarios.
 A novel mapping rule  was conceived for improving RM detection capability and for simplifying the channel estimation with no loss of accuracy.
 At the receiver, a layer-by-layer RM detection algorithm and an enhanced algorithm were invoked for single sequence detection.
They are based on the nested structure of RM sequences, which is discovered to reveal the deeper relationship between  their sub-sequences.
Then an iterative RM detection and channel estimation algorithm  was also designed for the case where multiple active users co-exist in the system.
From the simulation results, our proposed algorithms have  a significant advantage over existing ones in terms of detection performance and computational complexity.

\end{document}